\documentstyle[epsf]{article}

\def\hmpc{{\rm\, h^{-1}Mpc}}

\def\kpc{{\rm\,kpc}}

\def\msun{{\rm\,M_\odot}}

\def\'{^{\prime}}
\def\avrg#1{{\langle #1 \rangle}}

\def\hq{{\hat q}}

\def\eg{{\it e.g., }}
\def\ie{{\it i.e., }}
\def\etal{{\it et al. }}
 
\def\etc{{\it etc. }}

\def\half{{\textstyle{1\over2}}}

\def\p3m{P$^3$M}

\def\spose#1{\hbox to 0pt{#1\hss}}
\def\lta{\mathrel{\spose{\lower 3pt\hbox{$\mathchar"218$}}
     \raise 2.0pt\hbox{$\mathchar"13C$}}}
\def\gta{\mathrel{\spose{\lower 3pt\hbox{$\mathchar"218$}}
     \raise 2.0pt\hbox{$\mathchar"13E$}}}
\def\ge{\mathrel{\spose{\lower 3pt\hbox{$-$}}
     \raise 2.0pt\hbox{$\mathchar"13E$}}}
\def\le{\mathrel{\spose{\lower 3pt\hbox{$-$}}
     \raise 2.0pt\hbox{$\mathchar"13C$}}}
\def\simlt{\lta}

\begin{document}

\title{\bf Implications of the Background Radiation for 
Cosmic Structure Formation}

\author{J. Richard Bond}
\date{}

\maketitle

\centerline{Canadian Institute for Theoretical Astrophysics,}
\centerline{CIAR Cosmology Program}
\centerline{ University of Toronto, Toronto, ON M5S 1A7, Canada }

\begin{abstract}
The cosmic microwave background (CMB) is a remarkably distortionless
blackbody, and this strongly constrains the amount of energy that can
have been injected at high redshift, thereby limiting the role that
hydrodynamical amplification can have played in cosmic structure
formation. The current data on primary anisotropies (those calculated
using linear response theory) provide very strong support for the
gravitational instability theory and encouraging support that the
initial fluctuation spectrum was not far off the scale invariant form
that inflation (and defect) models prefer. By itself, the (low
resolution) 4-year DMR data allow relatively precise $\sigma_8$
normalization factors for density fluctuation spectra and rough
information on the large scale slope of the anisotropy power, thereby
focusing our attention on a relatively narrow set of viable
models. Useful formulae relating the DMR bandpower to $\sigma_8$ and
post-inflation scalar and tensor power spectra measures are
given. Smaller angle data (\eg SP94, SK94) are consistent with these
models, and will soon be powerful enough to strongly select among the
possibilities, although there remains much room for surprises.  In
spite of foregrounds, future high resolution experiments should be
able to allow precise determination of many combinations of the
cosmological parameters that define large scale structure formation
theories: mode (adiabatic/isocurvature, gravity wave content), shape
functions ($n_s(k),n_{is}(k), n_t(k)$), amplitudes ($\sigma_8 , {\cal
C}_2^{(T)}/{\cal C}_2^{(S)}$), and various mean energy densities $\{
\Omega,{\rm h}, \Omega_B , \Omega_\Lambda , z_{reh} , \Omega_{m\nu
}\}$. Secondary anisotropies arising from nonlinear structures will be
invaluable probes of shorter-distance aspects of structure formation
theories.
\end{abstract}

\vskip 20pt
\thanks{\centerline{in {\it
%\footnote{in {\it
The Evolution of the Universe}, pp. 199--223,} 
\centerline{ed. S. Gottlober \& G. Borner,
(Chichester: Wiley) (1997)}}

%\thanks{{\it CITA-95-28}, sissa: astro-ph/9512142}
\section{Basics of CMB Anisotropy} \label{intro}

We are in the golden age for cosmic background radiation research,
with signals unveiled by very high precision spectrum and angular
anisotropy experiments revealing much about how structure arose in the
Hubble patch in which we live. The main goal of theoretical anisotropy
research is to work out detailed predictions within a given cosmic
structure formation model of primary and secondary CMB temperature
fluctuations as a function of scale 
(\eg \cite{bh95,SilkScottWhiteannrev}).  {\it Primary anisotropies} are
those that we can calculate either fully with linear perturbation
theory, or, as in the case of cosmological defect models, with linear
response theory of nonlinear seed fluctuations.  Because of the
linearity, primary anisotropies are the simplest to predict and offer
the least ambiguous glimpse of the underlying fluctuations that define
the structure formation theory. With detailed high precision
observations, we expect to be able to use CMB anisotropies to measure
various cosmological parameters to remarkable accuracy 
(\eg  \cite{bcdes,knox,jkks96,map96ref,psifire96,cobrassamba96ref,bet97,zss97}).

Accompanying spectral distortions to the CMB that may be generated
during the evolution of nonlinear objects, there will be inevitable
{\it secondary} anisotropies that carry invaluable information about
the epochs that the relevant structures formed. Even if the
angle-averaged distortions are well below the level that absolute
spectrum experiments like COBE's Far Infrared Absolute
Spectrophotometer (FIRAS) probe \cite{fixsenTcmb}, it is certain 
that these secondary
anisotropies are accessible to experiments (\eg \cite{bh95}): 
the question is only for
what fraction of the sky do they rise above experimental noise and the
primary signal.

To relate observations of anisotropy to theory, statistical measures
quite familiar from their application to the galaxy distribution have
been widely used. Denote the radiation pattern as measured here and
now by the two-dimensional random field $\Delta T (\hq) $, where $-\hq
= (\theta , \phi)$ is the unit direction vector on the sky (and $\hq$
is the direction the photons are travelling in).  For CMB
anisotropies, it is natural to expand the radiation pattern in
spherical harmonics $Y_{\ell m} (\theta , \phi)$ and define an
ensemble-averaged angular power spectrum:
\begin{eqnarray}
&& {\Delta T \over T_c } (\hq ) = \sum_{\ell m} a_{\ell m}Y_{\ell m}
(\hq ) \, , \ {\cal C}_\ell \equiv \ell (\ell +1) 
\avrg{ a_{\ell m}^{*} a_{\ell m} }  /(2\pi) \ . \label{eq:pspec.1a}
\end{eqnarray}
At high $\ell$, ${\cal C}_\ell$ corresponds to the power in a
logarithmic waveband $d\ln (\ell )$. If the temperature pattern is
statistically isotropic, then $\avrg{ a_{\ell m}^{*} a_{\ell^\prime
m^\prime} } =0$ unless $\ell = \ell^\prime$, $m=m^\prime$. If the
initial fluctuations are Gaussian so is the primary CMB, hence ${\cal
C}_\ell$ is all that would be needed to characterize the anisotropy
statistics. Equivalently, the Gaussian patterns are completely
specified by the associated 2-point correlation function, $C(\theta )
\equiv \avrg{{\Delta T } (\hq ) {\Delta T } (\hq^\prime )}/T_c^2$,
where $\cos(\theta) = \hq \cdot \hq^\prime$. (If the statistics are
not Gaussian, then an infinite number of $N$-point correlation
functions are required to specify the statistical distribution.)
$C(\theta )$ and more generally the {\it rms} temperature anisotropies
associated with an $\ell$-space filter $W_{\ell}$ can be
expressed in terms of a ``logarithmic integral'' ${\cal I}[f_\ell ]$
of a function $f_\ell$:
\begin{eqnarray}
&&C(\theta ) = {\cal I} ({\cal C}_\ell P_\ell (\cos(\theta ))\, , , \quad 
\Big({\Delta T \over T}\Big)^2_{rms} (W_\ell ) \equiv {\cal I}\Big[
\overline{W}_{\ell}{\cal C}_{T\ell} \Big]\, , \\
&&  {\cal I}[f_\ell ]
\equiv \sum_\ell {(\ell+\textstyle{1\over2})\over \ell (\ell +1) }
f_\ell \, . \nonumber 
\end{eqnarray}
Even if an experiment has perfect resolution and
all-sky coverage, because the observed sky is just one realization
from the ensemble the derived ${\cal C}_\ell$ and $C(\theta )$ would
differ from the ensemble-averaged ones. This effect is called `cosmic
variance' and for example implies that, multipole by multipole, the
uncertainty is $\Delta {\cal C}_\ell = {\cal C}_\ell /(\ell
+\textstyle{1\over2} )^{1/2}$.

\begin{figure}
\vspace{-.5in}
\centerline{\epsfxsize=6.0in\epsfbox{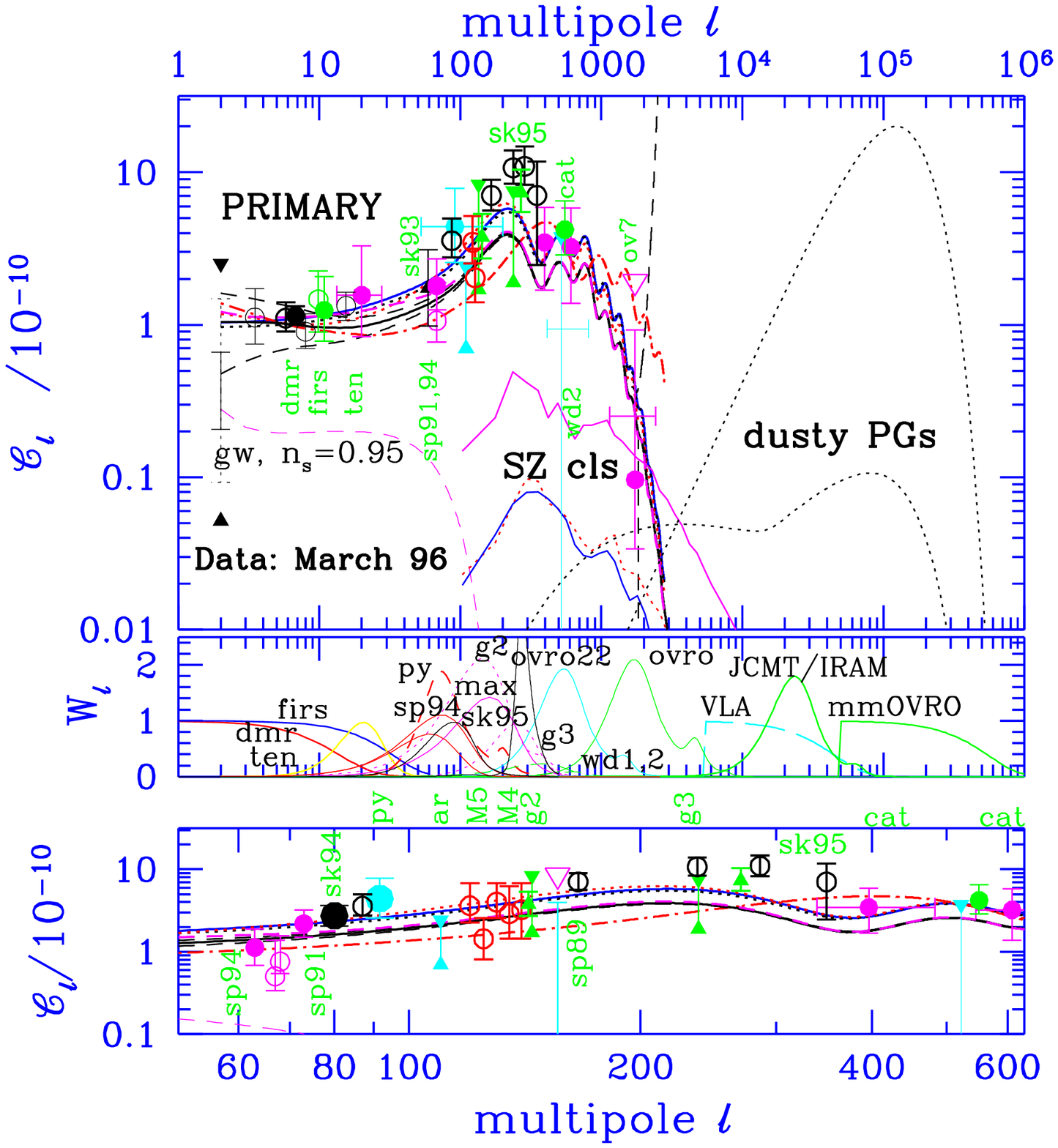}}
\vspace{-.3in}\
\centerline{ (caption next page) }
\label{fig:pow}
\end{figure}

Data from an anisotropy experiment are usually expressed in terms of
measurements $\overline{(\Delta T/T)}_p$ of the anisotropy in the
$p^{th}$ pixel and a pixel-pixel correlation matrix $C_{Dpp^\prime}$
giving the variance about the mean for the measurements. The signal
$({\Delta T/ T})_p$ can be expressed in terms of linear filters ${\cal
F}_{p,\ell m }$ acting on the multipole components, $a_{\ell m}$:
$\big(\Delta T /T \big)_p =\sum_{lm} {\cal F}_{p,\ell m } a_{\ell m}$,
where ${\cal F}_{p,\ell m }$ encodes the experimental beam and the
switching or modulation strategy that defines the temperature
difference. The former filters high $\ell$, the latter low $\ell$. A
given theory with power spectrum ${\cal
C}_{T\ell}$ has a pixel-pixel correlation matrix
\begin{eqnarray} 
C_{Tpp^\prime} &\equiv &
\avrg{\Big({\Delta T \over T}\Big)_p\Big({\Delta T \over
T}\Big)_{p^\prime}}  = {\cal I}\Big[ {\cal
C}_{T\ell}\  {4\pi \over
2\ell +1}  \sum_{m} {\cal F}_{p,\ell m } {\cal F}_{p^\prime ,\ell m
}^* \Big] \ .  \label{eq:pspec.ctpp}
\end{eqnarray}
We define the band-power of the experiment 
to be the anisotropy power across the average 
filter $\overline{W}_{\ell}$: 
\begin{eqnarray}
&& \avrg{{\cal C}_\ell}_{W}\equiv {\cal I}\Big[ \overline{W}_{\ell}{\cal
C}_{T\ell} \Big]/{\cal I}\Big[ \overline{W}_{\ell}\Big] = \Big({\Delta
T \over T}\Big)^2_{rms}(\overline{W}_\ell ) /{\cal I}\Big[
\overline{W}_{\ell}\Big] \, , \nonumber \\ 
&&\overline{W}_{\ell} \equiv
{1\over N_{pix} } \sum_{p=1}^{N_{pix}} {4\pi \over 2\ell +1} \sum_{m}
{\cal F}_{p,\ell m } {\cal F}_{p ,\ell m }^*\, . \label{eq:bandpow}
\end{eqnarray}
Usually the band-power is the quantity that can be most accurately
determined from the experimental data. Estimates of band-powers
derived for recent experiments (up to March 1996) are shown in
Fig.~\ref{fig:pow}. 
%(The Hubble parameter is ${\rm h} \equiv H_0/(100 \kms \mpc^{-1} )$.) 

 To determine band-powers for an experiment, a
local model of ${\cal C}_\ell$ is constructed, assumed to be valid
over the scale of the experiment's average filter $\overline{W}_\ell$.
A popular 2-parameter phenomenology has a  broad-band tilt
$\nu_{\Delta T}$ as well as a broad-band
power:
\begin{eqnarray}
{\cal C}_\ell &=& {\langle {\cal C}_\ell\rangle}_{\overline{W}} {{\cal
U}_\ell \ {\cal I}[
\overline{W}_{\ell}]\over 
{\cal I}[ \overline{W}_{\ell} {\cal U}_\ell ]}\, , \quad {\cal U}_\ell
\equiv {\Gamma\big( \ell +{\nu_{\Delta T} \over 2}\big) \Gamma(\ell
+2) \over \Gamma(\ell) \Gamma\big( \ell +2-{\nu_{\Delta T}
\over 2}\big)} \approx (\ell+\textstyle{1\over2})^{\nu_{\Delta T}}\ . 
\label{eq:CLpow}
\end{eqnarray}
As the data improves, a parameterized sequence of best-fit ${\cal
C}_\ell$'s will be preferable. 

Because there are so many detections now, Fig.~\ref{fig:pow} is split
into an upper and lower panel for clarity, with the upper giving an
overview, for experiments ranging from {\it dmr} at the smallest
$\ell$ to {\it ovro} at the highest $\ell$, and the lower panel
focussing on the crucial region of the first few peaks in ${\cal
C}_\ell$. 
Data points denote the maximum likelihood
values for the band-power, the error bars give the 16\% and 84\%
Bayesian probability values (corresponding to $\pm 1\sigma$ if the
probability distributions were Gaussian), and upper and lower
triangles denote 95\% confidence limits unless otherwise stated. The
horizontal location is at $\avrg{\ell}_W$ and the horizontal error
bars (where present) denote where the filters have fallen to
$e^{-0.5}$ of the maximum. The filters $\overline{W}_\ell$ for the
experiments are shown in the middle panel.

\begin{figure}
%\vspace{-.5in}
%\centerline{\epsfxsize=6.0in\epsfbox{dahlemCLdat.eps}}
%\vspace{-.3in}
\caption{ (previous page) {Band-power estimates derived for the 
anisotropy data up to March 1996. The lower panel is a closeup of the
first two `Doppler peaks'. The theoretical primary power spectra are
normalized to the 4-year {\it dmr}(53+90+31)(a+b) data. 
A ``standard'' $n_s$=1 $\Omega =1$ CDM
model with normal recombination, ${\rm h}=0.5$, $\Omega_B {\rm
h}^2=0.0125$  \protect{\cite{bbn}}, is the upper solid curve. 
It has $\sigma_8=1.20\pm
0.08$. An (almost indistinguishable) dotted curve has the same
parameters except that it includes a light neutrino with
$\Omega_{m\nu} =0.2$ (and $\Omega_{cdm}=0.75$). It has $\sigma_8
=0.83\pm 0.06$.  The upper dashed curve is a vacuum-dominated model
with a 13Gyr age and ${\rm h}=0.75$ ($\Omega_\Lambda =0.73$,
$\Omega_B=0.02$, $\Omega_{cdm}=0.24$). It has $\sigma_8=1.03\pm
0.07$. The model whose peak is shifted to high $\ell$ is an open CDM
cosmology \protect{\cite{bsouradeep96}} with the same 13 Gyr age, but now
$H_0=60$, and $\Omega_{tot}=0.33$ (and $\Omega_{cdm}=0.30$,
$\Omega_B=0.035$). It has $\sigma_8=0.44\pm 0.04$. 
By $H_0=70$, $\Omega_{tot}$ is down to 0.055 at
this age. The lower solid curve is a CDM model with reionization at
$z_{reh}=30$, and almost degenerate with it is a $n_s$=0.95
standard CDM model, whose gravity wave induced component 
is also shown. Power spectra of SZ maps
constructed using the peak-patch method \protect{\cite{bm3}} 
are shown for a $\sigma_8=1$ standard CDM model and a more realistic  
$\sigma_8$=0.7 hot/cold hybrid model (solid, $\Omega_{m\nu} $=0.3)
and a $n_s$=0.8 tilted CDM model. Although ${\cal C}^{(SZ)}_\ell$ may be
small, because the power for such non-Gaussian sources is concentrated
in hot or cold spots the signal is detectable. Spectra for a dusty
primeval galaxy model that satisfies the FIRAS constraint is also
shown, the larger (arbitrarily normalized) part a shot-noise effect
for galaxies with dust distributed over $10 \kpc$, the smaller a
continuous clustering contribution, including a nonlinear correction.
Average filter functions for a variety of experiments are shown in the
middle panel.  Error curves ($1\sigma$) on
the $z_{reh}=30$ model assume a homogeneously weighted all-sky survey
with a $10^\prime$ beam and a $6 \mu K$ per $10^\prime$ pixel noise,
which dominates at high $\ell$, while cosmic variance $\sim
\avrg{\ell}^{-1}$ dominates at low $\ell$; $\Delta \ln \ell =0.1$ was
used.  Current  band-powers broadly follow 
inflation-based expectations, but could include residual signals 
from systematic effects such as sidelobe 
contamination, Galactic
effects such as bremsstrahlung and dust, 
or secondary anisotropy signals.}}
\end{figure}

For the CMB data sets that have been obtained to date, including COBE,
it is possible to do complete Bayesian statistical analyses. To
determine the best error bars on the parameters of a target theory
with correlation matrix $C_{Tpp^\prime}$, a recommended method for
this analysis \cite{bdmr294,bh95,bj96} is to expand in signal-to-noise
eigenmodes, those linear
combinations of pixels which diagonalize the matrix
$C_n^{-1/2}C_TC_n^{-1/2}$, where the noise correlation matrix $C_n =
C_D+C_{res}$ consists of the pixel errors $C_D$ and the correlation of
any unwanted residuals $C_{res}$, whether of known origin such as
Galactic or extragalactic foregrounds or unknown extra residuals
within the data. This facilitates the many inversions of $C_n+C_T$
required to evaluate the likelihood function, and can also be a
powerful probe of unknown residuals contaminating the data. The $S/N$
mode expansion was used to get most of the bandpowers and their error
bars shown in Fig.~\ref{fig:pow}.

With uniform weighting and all-sky coverage, the
$S/N$-modes are just the independent $Re{(a_{\ell m})}$ and
$Im{(a_{\ell m})}$. 
The uniform noise assumption has been used recently to address the ultimate
accuracy that satellite experiments might achieve
 \cite{knox,tegmarkgpe,jkks96,cobrassamba96ref,bet97,zss97}, 
and is used here in Figs.~\ref{fig:pow} and \ref{fig:CLdiff} for that
purpose. The target power spectrum has ${\cal C}_\ell$ determined
within a $1\sigma$ deviation $\Delta {\cal C}_\ell $ given by
\begin{eqnarray}
&& \Delta {\cal C}_{T\ell} \approx { [({\cal C}_{T\ell} + {\cal
C}_{res,\ell}  + {\cal
C}_{D\ell} {\cal B}_\ell^{-2}] \over \sqrt{(\ell +\textstyle{1\over2})
f_{sky}} \, \sqrt{{\rm
cosh}(\Delta \ln \ell )[1+(\ell +\textstyle{1\over2}){\rm sinh}(\Delta
\ln \ell ) ]} } ,  \\
&&\quad {\cal
C}_{D\ell} \equiv {\ell (\ell +1) \over 2 \pi }\sigma_\nu^2 \,
. \nonumber 
\end{eqnarray}
If only a
fraction $f_{sky}$ of the sky is covered, then for high $\ell$, so
that the angular scale $\ell^{-1}$ is small compared with the patch
probed, the effective pixel number scales by $f_{sky}$.  
The errors are those
appropriate for logarithmic binning of width $\pm \textstyle{1\over2}
\Delta \ln \ell $ about $\ln \ell$, with $\textstyle{1\over2} \Delta
\ln \ell =0.05$. \footnote{This generally introduces smoothing
functions but these are nearly unity if $\Delta \ln \ell \ll 1$. If
$\Delta \ln \ell$ is so small as to encompass only one $\ell$ we
recover the usual $(\ell +\textstyle{1\over2})^{-1/2}$ cosmic variance
result.} The filter function associated with the beam is ${\cal
B}_\ell$, where 
\begin{eqnarray}
&& {\cal
B}_\ell = \exp\Big[-\half {(\ell +\half)^2 \over (\ell_s +\half)}\Big]
\quad {\rm and} \quad \ell_s \sim (0.425 \theta_{fwhm})^{-1}
\end{eqnarray}
for a Gaussian beam. It has been divided out to show that the
effective noise level picks up enormously above $\ell_s$.  The
parameter $\sigma_\nu$ is the error-per-pixel times $\theta_{pixel}$.

The lowest primary anisotropy curve in Fig.~\ref{fig:pow}, the
$z_{reh}=30$ model, has a set of one-sigma error bar curves (dotted)
on it associated with this uniform all-sky coverage error, at small
angles due to cosmic variance and at large due to pixel noise, with
the {\it fwhm} chosen to be $10^\prime$ ($\ell_s = 450$) and a noise
level of $6\mu K$ per $10^\prime$ pixel (and with $f_{sky}=1$). These
error curves are not even visible in the $\ell$ range of the lower
panel. These values are consistent with what might be expected from a very
high precision satellite experiment like COBRAS/SAMBA 
 \cite{map96ref,cobrassamba96ref}.

\begin{figure}
\epsfxsize=\hsize
\vspace{-.7in}
\centerline{\epsfxsize=5.0in\epsfbox{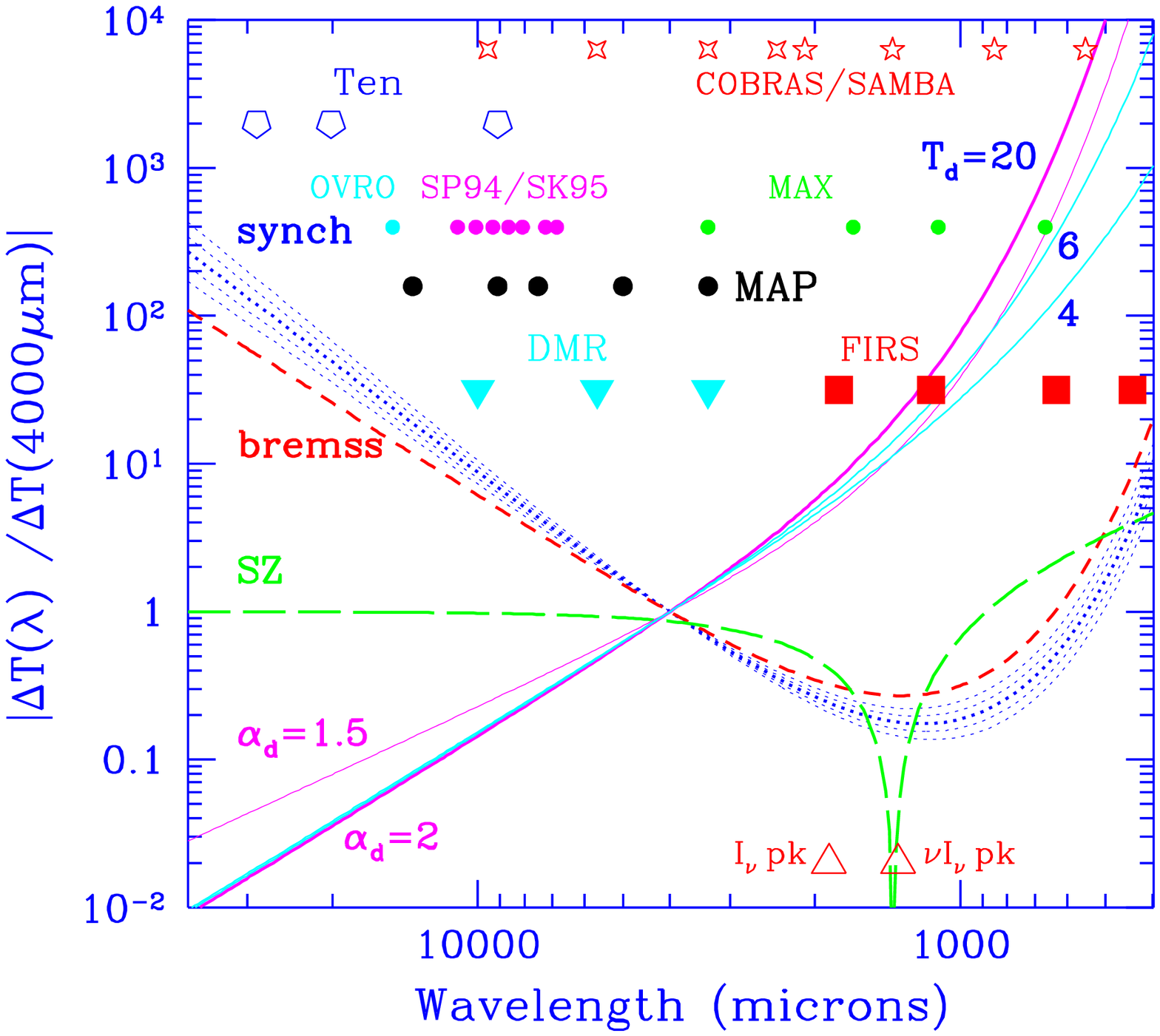}}
\vspace{-.4in}
\caption{ The flat spectrum in thermodynamic temperature predicted for
primary anisotropies is contrasted with the spectral signatures for
other sources of anisotropy (normalized at 4 mm): SZ anisotropies
(long-dashed, with a sign change at $1300~\mu {\rm m}$);
bremsstrahlung (short-dashed); synchrotron (dotted), with index
vvarying from $p_s =0.5$ to $0.9$ (intensity $I_\nu \propto \nu^{-p_s}$); 
and dust (with index $\alpha_d=2$ as
indicated by a two-temperature fit to COBE), both the usual Galactic
dust at 20~K (heavy solid) and dust at 6~K and 4~K (light solid lines,
which could represent a cold Galactic component or, \eg 30~K dust
radiating at redshift $\sim 5$); a shallower, less
physically-motivated, $\alpha_d=1.5$ dust opacity law for the 20K
grains is also shown, appropriate for the single-temperature COBE
fit. The frequency bands which various experiments probe are
indicated, in particular the proposed bands for the NASA (MAP) and ESA
(COBRAS/SAMBA) all-sky satellite experiments. There is a minimum of
the Galactic foregrounds at about 90 GHz, the highest frequency COBE
channel.}
\label{fig:dtspec} 
\end{figure}

What will limit this rosy picture is our ability to subtract
foregrounds. Ultimately, it will probably require a sophisticated
combination of spectral and angular information, and cross-correlation
with other datasets, such as X-ray and HI maps. With enough frequency
bands covered, the prospects for separation on the basis of spectrum
alone is good.  Figure~\ref{fig:dtspec} draws together the spectral
signatures of the different sources of anisotropy that are likely to
appear and compares them with the frequencies that various experiments
probe. Although the different signals are gratifyingly different, many
parameters must be fit, either pixel-by-pixel, or using as well the
different angular patterns that the signals will have. For example,
extragalactic radio sources will have synchrotron spectra and a
projected white-noise spectrum ${\cal C}_{res,\ell} \sim \ell^{2}$
like that shown in Fig.~\ref{fig:pow} for primeval galaxies; just as
for the primeval galaxies, there could also be clustering
contributions. The primeval galaxy frequency spectrum would be similar to that
of a cold dust component because of redshifting.  The angular power
spectra of Galactic bremsstrahlung and dust appear to obey ${\cal
C}_{res,\ell} \sim \ell^{-1}$, \ie fall to high $\ell$ faster than
scale invariance, apparently becoming small in the all important $\ell
\sim 100-500$ range, especially in the frequency range around 90
GHz \cite{gautier92,kogut95ff}. Complications will arise however, 
the most important being the
non-Gaussian nature of the residuals and the multicomponent nature of
the dust, in particular the possible presence of cold dust
\cite{Reachcolddu95,pugetfirb96}.

\section{CMB Distortions and Energetic Constraints} \label{energy}

In this section, I review the impact of spectrum observations on
structure formation issues. We know from FIRAS that the CMB is well
fit by a blackbody with $T\approx 2.728 \pm 0.004$~K over the region
from $5000 \mu {\rm m}$ to $500 \mu {\rm m}$ \cite{fixsenTcmb}, a
number compatible with the 1990 COBRA rocket experiment of Gush \etal
\cite{GushHalpernWishnow} covering the same band, and also with ground
based measurements at centimetre wavelengths --- although there is
still room for significant spectral distortion longward of 1 cm. We
must rely on indirect arguments based on primordial nucleosynthesis to
constrain exactly when this photon entropy in our Hubble patch came
into being, and whether this injection of energy would have a direct
impact on short-distance structure formation. Energy injection prior
to $z_{Pl} \approx 10^{6.9} \Big({\Omega_B {\rm h}^2 /
0.01}\Big)^{-0.39}$ is redistributed into a Planckian form: $z_{Pl}$
defines the redshift of the cosmic photosphere.  There have been
heroic efforts to explain the CMB as starlight processed through
exotic forms of dust (in particular long conducting needles) that
would have happened much later. The observables from these models have
never been fully worked out (see, \eg \cite{bh95}), but they are
severely challenged by the absence of spectral distortions and the
high degree of anisotropy required.

Between $z_{Pl}$ and $z_{BE} \approx 10^{5.6} \big({\Omega_B {\rm h}^2
/ 0.01}\big)^{-1/2}$ injected energy is redistributed into a
Bose-Einstein shape characterized by a chemical potential, which FIRAS
constrains to be  \cite{fixsenTcmb} 
$\vert \mu_\gamma \vert /T_\gamma < 0.9\times
10^{-4}$ (95\% \ CL), translating to a limit on energy of
${\delta E_{BE} / E_{cmb}} \lta 6.4 \times 10^{-5}$ in this redshift
range. Below $z_{y} \approx 10^{5} \big({\Omega_B {\rm h}^2 /
0.01}\big)^{-1/2}$ the Compton $y$-distortion formula holds, giving a
unique signature to distortions, negative for frequencies below $218$
GHz, positive above, and a stringent limit on the Compton-cooling
energy loss from hot gas, ${\delta E_{Compton\ cool} / E_{cmb}} = 4y <
6.0 \times 10^{-5}$ (95\% CL).  If there is no recombination, there is
a constraint from the $y$-distortion on how early reheating of the
Universe can have occurred: $z_{max, reh} \approx 10^{3.7}
\big({\Omega_B {\rm h} / 0.02}\big)^{-2/3} \Omega_{nr}^{1/3}$  
\cite{ZS69,BartlettStebbins,bh95}, 
but it is not very restrictive for
the low $\Omega_B$ favoured by standard Big Bang nucleosynthesis and
can be avoided if one can sustain a temperature of the cooling
electrons to be nearly the CMB temperature.

Compton cooling has been observed in more than two dozen massive clusters of
galaxies above the $5$-$\sigma$ level,
including one at redshift 0.545, which tells us that the CMB existed
by at least that redshift. With likely experimental sensitivity
increases, the SZ effect may eventually offer a more powerful probe of
the cluster distribution than $X$-ray observations do: instead of
being a projection of the square of the baryon density, it is a
projection of the electron pressure and the decrease in signal with
redshift is only a consequence of cluster evolution, not dimming by
distance.  Combining the SZ and $X$-ray observations is one of the
main paths to $H_0$ (and in principle $q_0$), but is so far more
confusing than enlightening, with values ranging from small (\eg
$H_0=38 \pm 17$ for Abell 2218) to large ($74\pm 29$ for COMA); the
hope is that a well-selected sample of clusters may help to reduce the
biases.

The integrated contribution of Compton-cooling from clusters and
groups is not expected to be large for models of structure formation
that reproduce the cluster X-ray temperature distribution
function. For example \cite{bm3}, for variants of 
adiabatic dark-matter dominated
$\Omega =1$ models with $\Omega_B \ll \Omega $, ${\rm h}=0.5$ and
nearly scale-invariant initial conditions, the estimated value depends
sensitively on $\sigma_8$, the linear amplitude of density fluctuations
on the cluster-scale $8 \hmpc$, and somewhat on the local curvature of
the density fluctuation spectrum on cluster-scales: with $\sigma_8
=0.7$, a hot/cold hybrid model with $\Omega_\nu =0.3$ gives
$\bar{y}\sim 0.3\times 10^{-6}(\Omega_{Beff}/0.05) $, a tilted CDM
model with $n_s=0.8$ gives $\sim 0.5\times
10^{-6}(\Omega_{Beff}/0.05)$, with a similar value obtained for a
$n_s=1$ model. Lowering $\sigma_8$ gives values well below $10^{-6}$,
but also not enough high temperature clusters; raising it gives too
many high $T_X$ clusters (raising $\sigma_8$ to unity still gives only
$\bar{y} \sim 1.6\times 10^{-6}(\Omega_{Beff}/0.05)$). Here
$\Omega_{Beff}$ takes into account the possible segregation of baryons
from mass in clusters, modifying the value to be used over the
primordial $\Omega_{B}$.  There is an even smaller effect associated
with nonlinear Thompson scattering from the hot gas in the moving
clusters.  Nonetheless, the non-Gaussian pattern of Compton-cooling
secondary anisotropies is accessible to experiment and will be a
foreground to remove in future CMB anisotropy experiments 
 \cite{cobrassamba96ref}.

The FIRAS limit on general secondary backgrounds (without a unique
signature like BE or $y$ distortions) is ${\delta E /
E_{cmb}}(500-5000 \mu {\rm m} ) <0.00025$ ($1\sigma$ CL). If
pregalactic dust, or dust in primeval galaxies, exists, it will absorb
higher frequency radiation (UV and optical) and down-shift it into the
infrared (\eg \cite{bch2}); combined with the redshift, 
a sub-mm background is expected
but, with FIRAS, is now quite strongly constrained.  The radiation
could be largely shortward of $500 \mu {\rm m} $: the peak in the $\nu
I_\nu$ curve occurs at $\lambda_{pk}\approx 96(1+z)(30{\rm K}/T_d)$
for $\alpha_d=2$ dust, which could be around $200 \mu {\rm m}$ if the
dust is hot (which seems reasonable) or the redshift of
bulge/elliptical formation is low.  The FIRAS constraint then applies
only to the tail of emission. There is a tentative identification of
a sub-mm background in the FIRAS data \cite{pugetfirb96} in the range
$\sim 200-1000~\mu {\rm m}$, with energy $\delta E/E_{cmb} \sim
10^{-3}$ longward of $\sim 400~\mu {\rm m}$, which partly mimics the
Galactic contribution (and could be partly due to cold high latitude
Galactic dust \cite{Reachcolddu95}).  There are also residuals after
source subtractions in the DIRBE data which could be interpreted as a
cosmological infrared background at shorter ($\sim 1-200~\mu {\rm m}$)
wavelengths at the $\delta E/E_{cmb} \sim 10^{-2}$ level
 \cite{dirbe95}. These constraints on energy injection should be
contrasted with plausible sources: For example, the nuclear energy output of
stars with efficiency $\epsilon_{nuc}$ radiating at redshift $z_*$
with an abundance $\Omega_*$ relative to the CMB is ${ E_* / E_{cmb}}
\approx 0.03 \big({\Omega_* {\rm h}^2 / 0.001}\big) \ {[5/ (1+z_*)]} \
[{\epsilon_{nuc}/ 0.004}]$.  Massive stars in the $10$--$100\msun$
range have $\epsilon_{nuc}\sim 0.004(M/100 \msun)^{0.5}$, saturating
at 0.004 above $>100 \msun$ (the Very Massive Object range). $E_*$ has
been used to constrain the role pregalactic black holes from VMO
precursors could have played as dark matter. The massive star $E_*$ is
also tied to the heavy elements $Z_{ej}M$ they eject in supernova
explosions. If the supernovae contribute a mean metal fraction $Z$ to
a gas of density $\Omega_{gas}$, ${ E_{preSN \, *} / E_{cmb}} \approx
0.0008 [{Z / 10^{-3}}] {[\Omega_{gas}{\rm h}^2 / 0.01]} [{Z_{ej}/
0.2}] \big({M / 20\msun }\big)^{0.5}$. Relaxation of the stringent
energetic constraints is possible if either the energy was not
reprocessed by dust (and so would reside in the near infrared where
the DIRBE constraints are not nearly as strong \cite{dirbe95}) or the
dust was so hot that even with redshift effects it was shortward of
$500 \mu m$.

As Ikeuchi and Ostriker originally emphasized, a predominantly
hydrodynamic explanation for cosmic structure development is a
perfectly reasonable extrapolation of known behaviour in the
interstellar medium to the pregalactic medium. However the Compton
cooling limit constrains the combination $f_{exp}\big({R_{exp} /
50\hmpc } \big)^2$ of filling factor $f_{exp}$ and bubble formation
scale $R_{exp}$ to be $\lta [(\Omega_{B} {\rm h}/0.02)
\Omega_{nr}^{1/2}]^{-1}$ (\eg \cite{bh95}). Further, if supernova
explosions were responsible for energy injection, one expects that the
presupernova light radiated would be much in excess of the explosive
energy (more than a hundred-fold), which would lead to much stronger
restrictions on the model; and if the supernova debris is
metal-enriched, the allowed amount of metals poses an even stronger
constraint.  (One may also argue that the specific tapestry that we
see is too close to what straightforward gravitational instability
predicts to warrant consideration of a purely hydrodynamical model;
\ie that the explosive effects might be largely masked by subsequent
gravitational instability. What does seem inevitable is that there
will be a more limited local hydrodynamics role around collapsed
objects.)

\section{Theoretical Issues and Sample Primary Power Spectra; COMBA}
\label{sechomtrans}

       The development of spectral distortions or angular anisotropies
in the microwave background is described by radiative transfer
equations for the photon distribution function, which are coupled to
Einstein's equations for the gravitational field and to the
hydrodynamic and transport equations for the other types of matter
present. The primary spectra are calculated by solving for each mode
$\in \{$adiabatic scalar, isocurvature scalar, vector or
tensor$\}$ the linearized Boltzmann transport equation for photons
(including polarization) and relativistic or light neutrinos, coupled
to the equations of motion for baryons and cold dark matter, and the
perturbed gravitational metric equations, possibly in the presence of
vacuum energy or mean curvature. This is a well developed art the
techniques for which have been described elsewhere 
(\eg \cite{bh95}) 
and will not be elaborated upon here. 
In homage to the high precision future that CMB experimentalists will
provide for us, a large consortium of theorists who have developed
computer codes to attack this problem fully or in various
fast-computation approximations have combined under the acronym COMBA
to deliver accurate validated calculations of ${\cal C}_\ell$'s to the
cosmological community \cite{comba95}. 

\begin{figure}
\vspace{-0.4in}
\centerline{\epsfxsize=5.0in\epsfbox{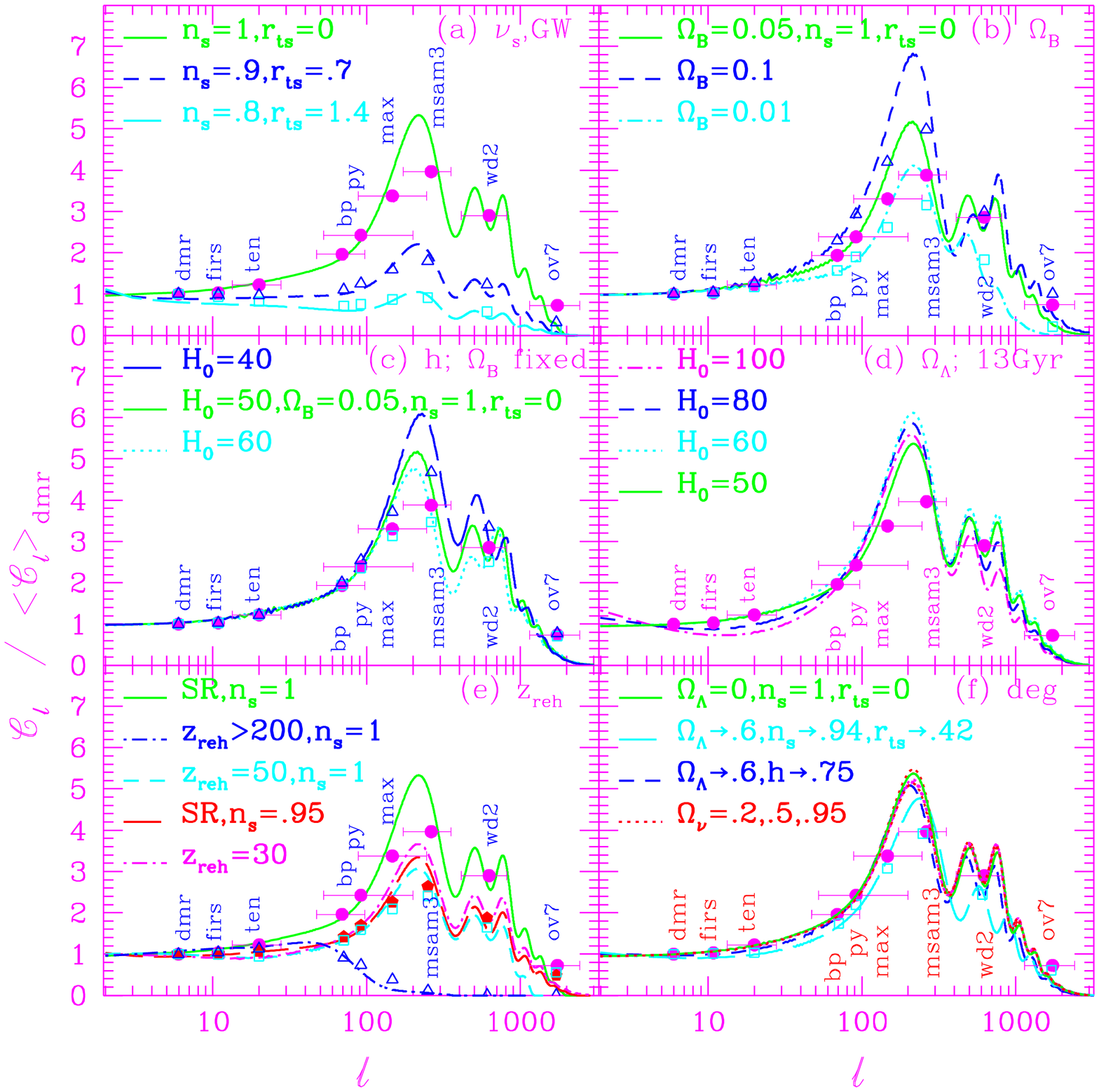}}
\vspace{-.2in}
%\centerline{ (caption next page)} 
%\label{fig:CLth6}
%\end{figure}
%\begin{figure}
%\vspace{-0.4in}
%\centerline{\epsfxsize=5.0in\epsfbox{dahlem_CL6.eps}}
%\vspace{-.2in}
\caption{ (previous page) Spectra for a variety of
inflation-inspired models, normalized to the COBE band-power.
Theoretical band-powers for various experimental configurations are
placed at $\avrg{\ell}_{W}$, horizontal error bars extend to the
$e^{-1/2}\overline{W}_{max}$ points. Unless otherwise indicated,
$\Omega_B{\rm h}^2$=0.0125, ${\rm h}$=0.5, $n_s$=1; when the gravity
wave contribution is nonzero, $\nu_t$=$\nu_s$ is assumed 
(and $r_{ts}\equiv{\cal C}_2^{(T)}/ {\cal C}_2^{(S)}$ is $\sim -7 \nu_t$). The
untilted $n_s$=1, $r_{ts}$=0 model is repeated in each panel (solid
line). (a) CDM models with variable tilt $n_s$. (b) $n_s$=1 models
with $\Omega_B{\rm h}^2$ changed, ${\rm h}$ fixed. (c) $n_s$=1 models,
with $\Omega_B{\rm h}^2$ changed, $\Omega_B$ fixed.  (d) $n_s$=1
models with fixed age, 13 Gyr, but variable $H_0$ and $\Omega_\Lambda
=1-\Omega_{cdm}-\Omega_{B}$ (.92,.79,.43,0 for 100,80,60,50).  (e) CDM
models with very early reionization at $z_{reh} \gta 150$ (equivalent
to `no recombination'), and later reionization at $z_{reh}$=30,50 are
contrasted with standard recombination (SR).  The $z_{reh}$=50
spectrum is close to the $n_s$=0.95 spectrum with SR (thin,
dot-dashed): the moderate suppression if $20 \lta z_{reh} \lta 150$
can be partially mimicked by decreasing $n_s$ or increasing ${\rm
h}$. (f) Sample cosmologies with nearly degenerate spectra and
band-powers. Dashed curve: increasing $\Omega_{\Lambda}$ is
compensated by increasing ${\rm h}$.  Dot-dashed curve: tilting to
$n_s$=0.94 ($\tilde{r}_{ts}$=0.42) is compensated by increasing
$\Omega_{\Lambda}$ to 0.6. The dotted hot/cold model curves
\protect{\cite{lithwick95}} (with
$\Omega_{\nu}$ indicated) are nearly identical to the
standard CDM one, but even these few percent differences can be
distinguished in principle by satellite all-sky experiments with
currently available detector technology.}
\label{fig:CLth6}
\end{figure}

I now sketch the ${\cal C}_\ell$ terrain in inflation-inspired models
as the parameters defining the structure formation model are
varied. Samples are shown in Figs.~\ref{fig:pow},\ref{fig:CLth6}. The
``standard'' scale invariant adiabatic CDM model ($\Omega =1$,
$n_s=1$, ${\rm h}=0.5$, $\Omega_B=0.05$) with normal recombination
shown in Fig.~\ref{fig:pow} and repeated in each of the panels of
Fig.~\ref{fig:CLth6} illustrates the typical form: the Sachs-Wolfe
effect arising from gravitational potential fluctuations dominating at
low $\ell$, followed by rises and falls in the first and subsequent
Doppler (or acoustic) peaks, arising from a combination of photon
compression and rarefaction and electron flow at photon decoupling,
with an overall decline due to destructive interference across the
photon decoupling surface and damping by shear viscosity in the photon
plus baryon fluid. A CDM model with very early reionization
(at $z> 200$) shows no Doppler peaks, a result of
destructive interference from forward and backward flows across the
decoupling region, illustrating that the ``short-wavelength'' part of
the density power spectrum can have a dramatic effect upon ${\cal
C}_\ell$, since it determines how copious UV production from early
stars was. Lower redshifts of reionization still maintain a Doppler
peak, but are suppressed relative to the standard CDM case (as
illustrated by the $z_{reh} \sim 30$ model in Fig.~\ref{fig:pow} and 
in Fig.~\ref{fig:CLth6}(e)).

 Figs.~\ref{fig:pow},\ref{fig:CLth6}
include adiabatic scalar and tensor contributions. The relative
magnitude of each is characterized by either the ratio of the
quadrupole powers, $r_{ts}={\cal C}_2^{(T)}/{\cal C}_2^{(S)}$, or the
ratio of the {\it dmr} band-powers $\tilde{r}_{ts}=\avrg{{\cal
C}_\ell^{(T)}}_{dmr}/\avrg{{\cal C}_\ell^{(S)}}_{dmr}$. For the scale
invariant cases, $r_{ts}$ is taken to vanish.

A simple variant of CDM-like models is to tilt the initial spectrum.
The scalar tilt $\nu_s = n_s-1$ is defined in terms of the index
$n_s$, which is one for scale invariant adiabatic fluctuations. There
is a corresponding tilt which characterizes the initial spectrum of
gravitational waves which induce primary tensor anisotropies, $\nu_t =
n_t+3$, where $n_t$ is $-3$ for a scale invariant spectrum. Inflation
models give $\nu_t < 0$ and usually give $\nu_s < 0$. For small tensor tilts,
$r_{ts}\approx -6.9 \nu_t$ and $r_{ts}\approx 1.3 \tilde{r}_{ts}$ are
expected (with corrections given by eq.~\ref{eq:PGPzeta}). For a
reasonably large class of inflation models $\nu_t \approx \nu_s$, but
in some popular inflation models $\nu_t$ may be nearly zero even
though $\nu_s$ is not.  Figs.~\ref{fig:pow} and \ref{fig:CLth6}(a)
show ${\cal C}_\ell^{(S)}+{\cal C}_\ell^{(T)}$
derived for tilted cases when $\nu_t = \nu_s$ is assumed to
hold. Fig.~\ref{fig:pow} shows explicitly the contribution that ${\cal
C}_\ell^{(T)}$ makes in one example.  The tilt indices, especially
$\nu_s$, can also be complex functions of $k$ in inflation models in
which scale invariance is radically broken (\eg \cite{sbb}), 
in which case the ${\cal
C}_\ell$ reflect the added complexity. 

Spectra for hot/cold hybrid models with a light massive neutrino look
quite similar to those for CDM only, as Fig.~\ref{fig:CLth6}(f) 
shows \cite{MaBert,dodelsonmnu95,lithwick95}. This is true even 
for pure hot dark matter
models \cite{be84}.  

The dotted ${\cal C}_\ell$ in Fig.~\ref{fig:pow} also has a flat
initial spectrum, but has a large nonzero cosmological constant in
order to have a high $H_0$, in better accord with most observational
determinations. As one goes from $\ell$=2 to $\ell$ =3 and above there
is first a drop in ${\cal C}_\ell$  
\cite{kofman85}, 
a consequence of
the time dependence of the gravitational potential fluctuations (the
integrated Sachs-Wolfe effect).  Other nonzero $\Lambda$ examples are
given in Fig.~\ref{fig:CLth6}(d).

Open models like the $H_0=60$ one shown in Fig.~\ref{fig:pow} have a
nontrivial late-time integrated Sachs-Wolfe effect, like the
$\Omega_\Lambda \ne 0$ models do, but there is also a direct effect of
the curvature on the mode function evolution, which serves to focus
the structure to a smaller angular scale ($\sim \Omega_{nr}^{1/2}$)
than in the $\Omega_{curv} \equiv 1 -\Omega_{tot} =0$ case. Of course
whatever mechanism generated the ultra-large-scale mean curvature may
well have had associated with it strong fluctuations on observable
scales, so much so that this is an argument against large mean
curvature because of the absence of such effects in the CMB. Even if
the background curvature is determined by an entirely different
mechanism, it should influence the fluctuation generation
mechanism. An open issue in open models has always been what is a
natural shape for the spectrum for $k$ near $d_{curv}^{-1}\equiv H_0
\Omega_{curv}^{1/2}$.  Power laws in $kd_{curv}$,
$\sqrt{(kd_{curv})^2-1}$ \etc have often been adopted but if the
fluctuation generation mechanism is quantum noise in inflation, there
is a natural adiabatic spectrum expected which is a simple
generalization of the nearly scale invariant spectra of
$\Omega_{curv}=0$ inflation models
\cite{ratrapjep94,kamionkowski94,bsouradeep96}.  Fortunately for $\ell
\gg 2 \Omega_{nr}^{-1} \Omega_{curv}^{1/2}$ this issue is not a
factor, so that intermediate and small angle predictions are
relatively unambiguous.

  Inflation-based models with isocurvature rather than adiabatic 
initial conditions
are strongly ruled out by the CMB data if they are nearly scale
invariant \cite{eb86,eb87}, but could contribute at a
subdominant level to the adiabatic fluctuations. Even allowing
for arbitrarily broken scale invariance in the initial fluctuation
spectra, the allowed region for pure isocurvature baryon or CDM 
models has been shrinking fast as the data has improved.

Defect models have (knot-like or string-like) localized topological
field configurations acting as isocurvature seed perturbations to
drive the growth of fluctuations in the total mass density. On large
angular scales the defect models lead to a similar nearly
scale-invariant spectrum for ${\cal C}_\ell$ as for inflation-inspired
adiabatic perturbation models
\cite{COBEstring2,COBEtexture,texturecouldson94,crittendenturok95,allenstring97,penturokseljak97}. 
On smaller scales, the spectra are
sufficiently different from adiabatic inflation-inspired spectra to
sharply test these competing pictures of cosmic structure formation in
the next generation of CMB experiments. The non-Gaussian nature of
defect-induced anisotropies also adds another point of differentiation
among the models.

The angular power spectra generally differ substantially as a function
of multipole $\ell$ so that even with the experimental results we
expect in the very near future major swaths of cosmological parameter
space can be ruled out. However, different combinations of the
parameters $n_s, r_{ts}, \Omega_B, \Omega_{\Lambda}$ \etc can lead to
nearly identical spectra \cite{bcdes}, as Fig.~\ref{fig:CLth6}(f)
illustrates. Superposed upon the spectra in Fig.~\ref{fig:CLth6} are
theoretical band-powers derived for a variety of anisotropy
experiments. Fig.~\ref{fig:CLth6} also shows 10\% 1-sigma error bars:
COBE achieved 14\% errors with 4 years of data; to
achieve this with smaller angle experiments one needs to have about
the same number of pixels as COBE, but scaled to the beam size hence
covering a smaller region of the Universe.  So far none of the smaller
angle data sets have the 650 or so {\it fwhm}-sized pixels COBE
does, but this is the stage we are now entering \cite{sk95}.   

\begin{figure}
\centerline{\epsfxsize=5.0in\epsfbox{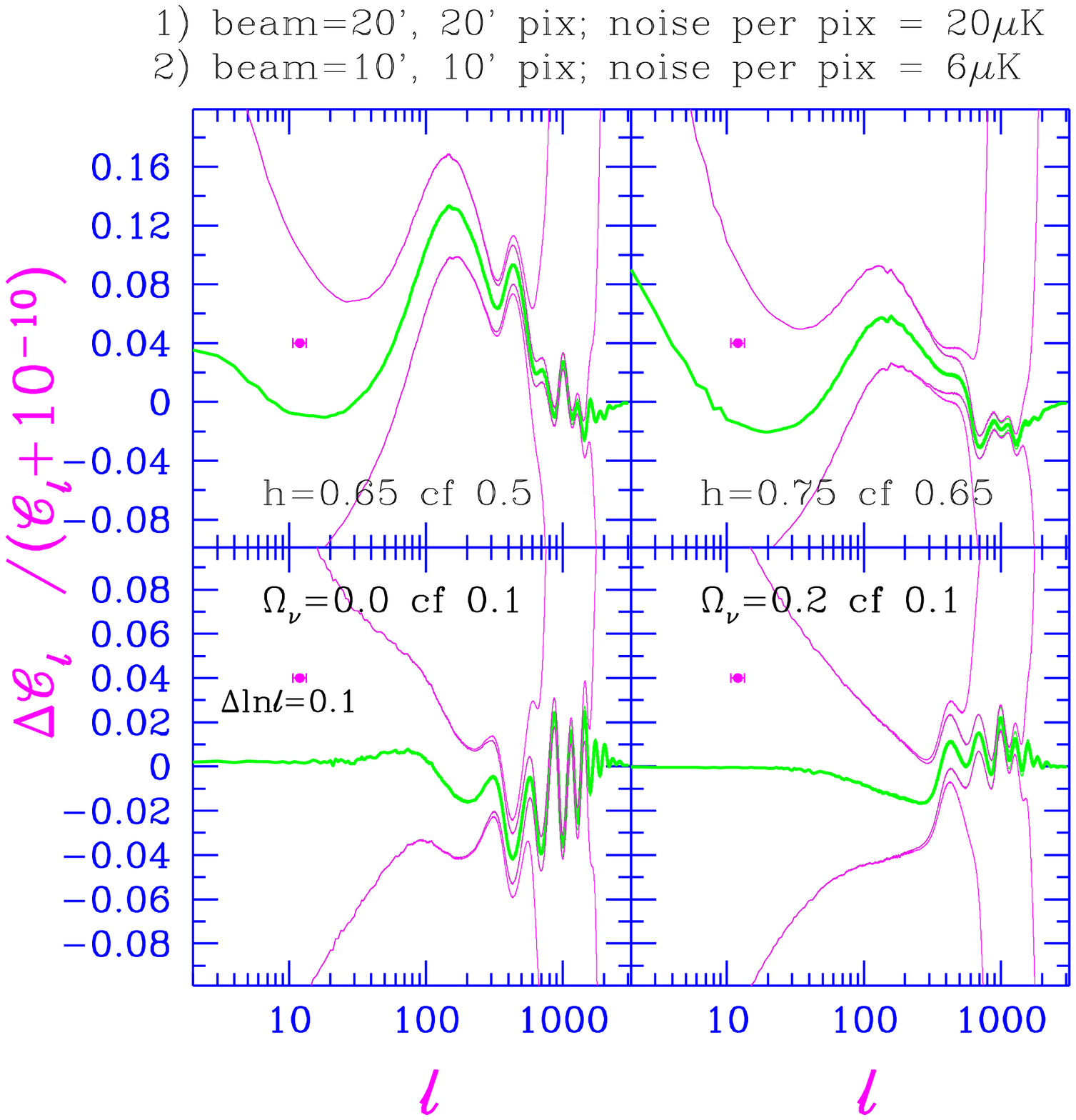}}
\vspace{-.2in}
\caption{This shows the ability of satellites to
measure cosmic parameters to high accuracy. The relative difference of
the power spectrum in question from a comparison spectrum (both
normalized to the 4-year {\it dmr} (53+90+31)A+B COBE maps) are shown
so that the few percent deviations can be clearly seen over the entire
$\ell$ range. The lighter lines are $1-sigma$ error bars for all-sky
coverage (averaged over the smoothing width shown, with
$\textstyle{1\over2}\Delta \ln \ell=0.05$) and include
cosmic variance (dominant at low $\ell$) and pixel noise at $20\mu
{\rm K}$ or $6\mu
{\rm K}$ (dominant at high $\ell$), with the very rapid growth
relative to the theory curve at high $\ell$ coming from the finite
beam-size (with the {\it fwhm} indicated, corresponding to a Gaussian
filter in multipole space of $\ell_s=404$ and $\ell_s=809$
respectively). The first choice corresponds to the NASA satellite
experiment MAP, the second choice to the ESA mission
COBRAS/SAMBA, assuming the entire sky is usable (errors scaling
$\propto f_{sky}^{-1/2}$).  The ultimate accuracy achievable will
depend upon the decontamination of the primary signal of non-Gaussian
Galactic synchrotron, bremsstrahlung and dust signals. The models
shown all have a uniform age of 13 Gyr, $\Omega_{cdm}+\Omega_{m\nu}+
\Omega_\Lambda +\Omega_B=1$, $\Omega_B{\rm h}^2 =0.0125$, $n_s=1$ and
no gravity wave contribution. Notice the scale change for the hot/cold
model panels. (One species of massive neutrino was adopted for these
two cases.) }
\label{fig:CLdiff}
\end{figure}

As noted before, even if there were idealized perfect all-sky coverage
with noise-free versions of the experiments of Fig.~\ref{fig:CLth6},
there would still be cosmic variance errors on the band-powers, but
these are $\sim \avrg{\ell}^{-1}$ 
\cite{capripow}, 
much smaller than the size of the points.  Thus it appears that by
using (perfect) CMB experiments which are sensitive to a wide range of
angular scales, we can distinguish even among the nearly
degenerate theoretical models shown, and be able to measure the
parameters that define the variations in these models.  A closeup view
of examples of how we can measure very fine differences in models is
shown in Fig.~\ref{fig:CLdiff}, using the detector sensitivities and
long observing times that satellite experiments
now currently feasible can achieve \cite{map96ref,cobrassamba96ref}.

\section{Secondary Anisotropy Sources} \label{secondary} 

\noindent
{\bf Reionization and Primary CMB Anisotropies:} An important issue
associated with the early energy injection described in
\S~\ref{energy} is what impact it will have on the primary
anisotropies of the CMB.  The physical processes important in the
recombination of the primeval plasma have been well understood since
shortly after the discovery of the CMB.  The comoving width of the
region over which decoupling takes place if there is normal
recombination is only $(5-10) \Omega_{nr}^{-1/2}\hmpc$, where
$\Omega_{nr}$ is the density in non-relativistic particles (CDM,
baryons); the viscous damping scale of the photon-baryon fluid prior
to recombination is slightly less. The associated angular scale can be
characterized by a multipole number, $\ell \sim 1000$, above which
anisotropies are strongly damped (Fig.~\ref{fig:pow}). The associated
natural `coherence' angle, $\sim 10^\prime$, defines which experiments
are most useful to do if we wish to probe the moment when the photons
were first released to freely propagate from their point of origin to
us, without much further modification, apart from some gravitational
redshifts, some lensing, and possibly some scattering from hot gas.

The main effect that reionization of the Universe has on anisotropies
is to lower their amplitude by a factor $\Delta T/T \propto
e^{-\zeta_C}$, where $\zeta_C$ is the optical depth to Thompson
scattering. If $z_{reh}$ is the reionization redshift and $z_{\zeta_C
= 1}\approx 10^{2.1} \big({\Omega_B {\rm h}/ 0.02 }\big)^{-2/3} \,
\Omega_{nr}^{1/3}$ is the redshift one would need to reionize by to
get a Thompson depth of unity, then $\zeta_C = [(1+z)/(1+z_{\zeta_C =
1})]^{3/2}$. With standard Big Bang nucleosynthesis values for
$\Omega_B$, getting $z_{\zeta_C = 1}$ much below 100 seems unlikely.
$z_{reh}$ is presumably the redshift by which the first nonlinear
objects form in sufficient abundance to allow enough massive star
formation to occur to cause pregalactic HII regions to overlap, a
quantity largely determined by the short-distance density fluctuation
power in the structure formation theory in question, but subject to
many uncertainties: the entities which form may well be rather fragile
with a small binding energy, easily disrupted by the massive stars
they generate; on the other hand, the amount of nonlinear gas could be
amplified by the explosion of such stars sweeping up shells of gas far
from the parent object.  Thus $z_{reh}$ depends upon how rare the
ionization-generators can be. For inflation-based CDM models and
variants with light massive neutrinos or nonzero $\Lambda$, $z_{reh}$
values ranging from 5 to 60 seem plausible, hence $\zeta_C$ ranges
from negligible to substantial, $ \lta 1/3$, but not so large as to fully
erase anisotropies, which  would occur on scales below
$\tau_{dec}/\pi$, where $\tau_{dec}$ is the horizon scale at photon
decoupling, corresponding to $\ell \sim 100$ for the ``standard'' CDM
model. See Figs.~\ref{fig:pow} and \ref{fig:CLth6}(e) for $z_{reh}\sim
30,50$ examples.

  In isocurvature baryon models with (nearly) white noise initial
conditions popular in the late seventies, the first objects collapse
at $z \sim 300$, making reionization easy, and, indeed,
expected. Thus, large viscous damping is expected. However, one can
still have a peak in ${\cal C}_\ell$ at $\ell \sim 200$ for open
models, essentially because the modified angle-distance relation in
curved universes shifts power to higher $\ell$.  Early ionization
seems plausible, but by no means certain, in models in which there are
isocurvature seeds, such as in texture models \cite{texturecouldson94}.

\noindent
{\bf Reionization and Quadratic Nonlinearities in Thomson Scattering:}
These can sometimes dominate over the first-order anisotropies if the
latter are strongly damped and there is early ionization. Even if
there is early reionization in nearly scale invariant models, there is
generally not sufficient power on small length scales for this
Vishniac effect \cite{Vishniac87} to be important. Thus it can usually
be ignored
in inflation-based models. This is not so for isocurvature baryon
models \cite{eb87} in which the initial spectral index $n_{is}$, 
considered a free
parameter, is between --1 and 0 on phenomenological grounds. Such a
steeply rising spectrum implies short-distance effects are very
important, and give predicted sizable signals in \eg the {\it ovro} and
{\it VLA} window of Fig.~\ref{fig:pow}.

\noindent
{\bf The Rees-Sciama Effect:} In flat $\Omega_{nr}=1$ models, the
gravitational potential fluctuations have constant amplitude in the
linear phase of evolution. Weak (or strong) nonlinear evolution
induces time dependence in $\Phi_N$ which induces anisotropy by an
integrated Sachs-Wolfe effect. For inflation-inspired models this
turns out to be a very small correction and can be largely ignored.

\noindent
{\bf The Influence of Weak Gravitational Lensing on the CMB:} Another
nonlinear effect (on the distribution function) is gravitational
lensing which bends, focusses and defocusses the CMB photons as they
propagate from decoupling through the clumpy medium to us. Given the
difficulties that astronomers have had detecting lensing, with the
best observations coming from clusters of galaxies, it may seem
obvious that the effect on the $\sim 10^\prime$ coherence scale
typical for primary CMB anisotropies is likely be quite small; and
this is what (most of) the people who have investigated the effect
have found. Lensing conserves the total angular power, it just
rearranges it, by smoothing the Doppler peaks. The typical range in
$\ell$ over which the power is spread in $\Delta \ell /\ell $ is
basically the weak-lensing shear, about 10\% to 20\% or so at a few
arcminutes, depending upon the model \cite{seljak94lens}; this is in
agreement with the levels estimated by people advocating using the
influence of weak-lensing on the ellipticities of faint galaxy images
to determine the mass density power spectrum \eg \cite{kaiser92lens}.

\noindent
{\bf Sunyaev-Zeldovich Fluctuations and the Moving Cluster Effect:}
The SZ effect has been discussed in \S~\ref{energy}.  The moving
cluster effect is the nonlinear Thomson scattering of the CMB photons
from plasma confined to clusters, moving with it. It is predicted to
be quite a bit smaller than the SZ anisotropy level, and as can be
seen from Fig.~\ref{fig:pow} the contribution of SZ from clusters is
small relative to the primary anisotropy power \cite{bm3}. However, the
distribution is non-Gaussian, concentrated in the pressure-peaks in
the medium, especially the clusters. This means a search for the
ambient SZ-effect (where the SZ sources are not known beforehand)
could be very promising \cite{cobrassamba96ref}.

\noindent
{\bf Starbursting Primeval Galaxies:} As discussed in \S~\ref{energy}
on CMB distortions and energetic constraints, 
the prospects for seeing these with sub-mm telescopes probing tens of
arcseconds ({\it e.g.} SCUBA filter in Fig.~\ref{fig:pow}) seem quite
good \cite{bch2,bh95}. 

\section{Anisotropy Experiments and their Band-Powers }
 \label{obspspec}

The importance of the large-angle COBE {\it dmr}
\cite{dmr12,dmr4} detection for testing theories of cosmic 
structure formation can
hardly be overstated. Smaller angle experiments are now also achieving
results that can be combined with COBE to constraint model
parameters. In this section, I review the experiments.

\begin{figure}
\vspace{-.3in}
\centerline{\epsfxsize=6.0in\epsfbox{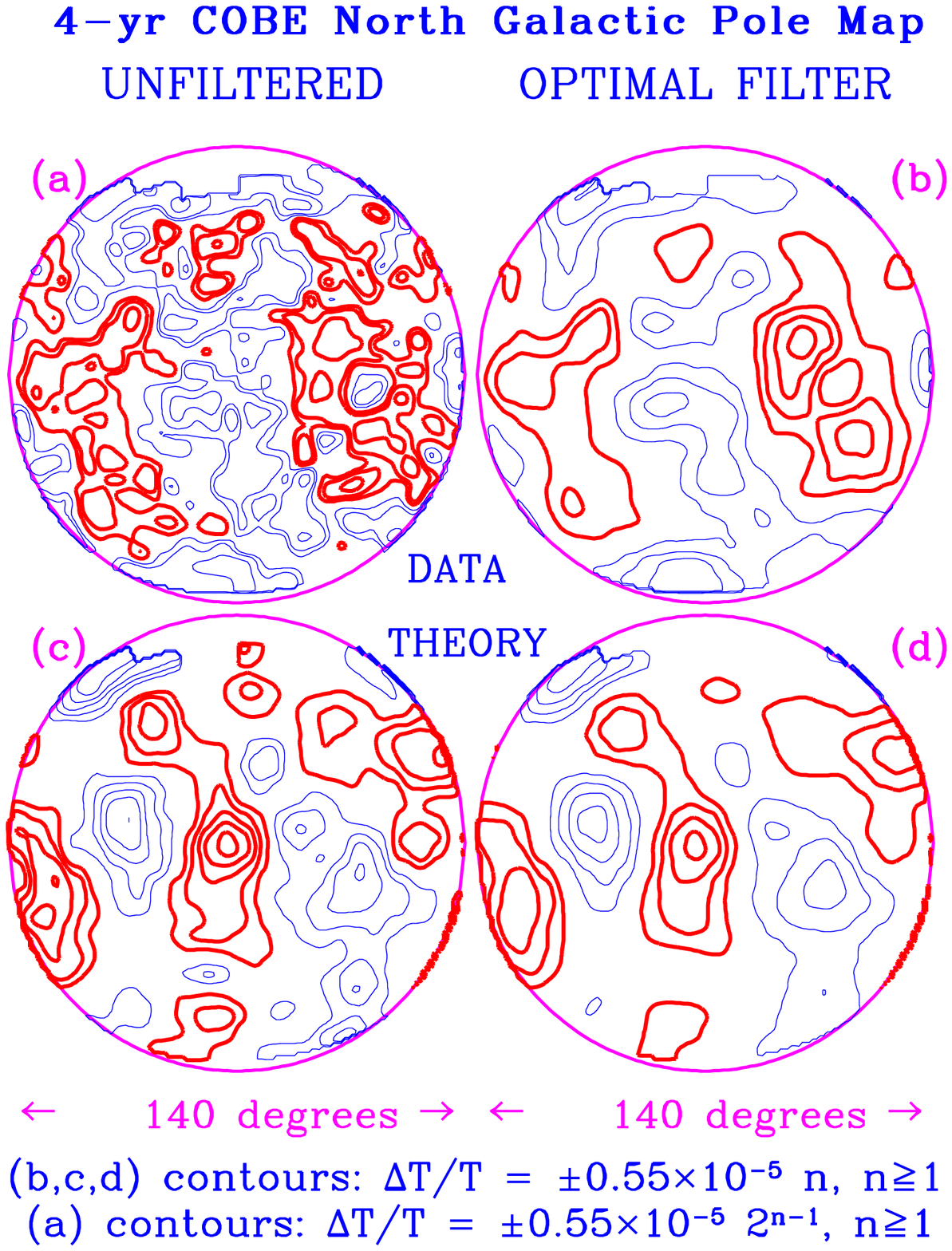}}
\vspace{-.2in}
\caption{ $140^\circ$ diameter maps centered on the North
Galactic Pole are shown for a realization of a CDM ${\cal
C}_\ell$--spectrum convolved with the {\it dmr} beam in (c). No noise
has been added. This is how the primary sky would appear in a $n_s=1$
CDM Universe with $\sigma_8 =1.2$ (or in a $\Omega_{m\nu}=0.2$
hot/cold universe with $\sigma_8 =0.8$), the most likely values for
the {\it dmr} data. 
This is contrasted with the 4-year {\it dmr} (53+90+31)a+b
data map shown in (a) and the map after the data has undergone optimal
signal-to-noise filtering in (b) (using the same ${\cal C}_\ell$-shape
and amplitude for the filter). The statistically significant features
are also seen in {\it each} of the {\it dmr} channel maps after
optimal filtering (which preferentially removes high angular
frequencies, more so for noisier maps). Thus, to compare, (d) shows
the theoretical realization after passing (c) through the same optimal
filter used for (b); the average, dipole and quadrupole of the full
$\vert b \vert > 20^\circ$ sky were also removed, an effective low
$\ell$-filter -- if they stay in, the maps look similar to the
unfiltered theory maps except small scale smoothing leads to loss of
the higher contour levels. Note that the contours are linearly spaced
at $\pm 15n~\mu {\rm K}$ for all but (a), for which the spacings are
$\pm 15,30,60,120~\mu {\rm K}$.  The maps have been smoothed by an
additional $1.66^\circ$ Gaussian filter.}
\label{fig:dmr4poles}
\end{figure}

The COBE data stream gives temperature differences in pixel pairs
separated by $60^\circ$, but an inversion yields 6 maps, for the $A$
and $B$ channels of the 3 frequencies shown in Fig.~\ref{fig:dtspec},
31, 53 and 90 GHz.  Figure~\ref{fig:dmr4poles} compares an optimally
filtered map of COBE's data with what a scale invariant $\Omega=1$
dark matter dominated model would look like under the same
filtering. 

 The $\ell=2$ power uses the 4-year quadrupole value
\cite{dmr4,dmr4k}, determined from high Galactic latitude data. It is
the multipole most likely to have a residual Galactic signal
contaminating it, possibly destructively, and the ``systematic''
error, the dashed addition to the statistical error bar (solid),
reflects this. The two heavy points at $\ell \sim 7$ are
band-powers derived for the 4-year {\it dmr} 53+90+31 GHz ``$A$+$B$''
maps \cite{bdmr294,bh95,bj96}, the
solid point assuming a $\nu_{\Delta T}=0$ spectrum, the open
marginalizing over all possible $\nu_{\Delta T}$, where $\nu_{\Delta
T}$ is defined by eq.(\ref{eq:CLpow}).  

The two points at $\ell \sim 10$ are for the {\it firs} map \cite{firsdmr},
solid with the restriction $\nu_{\Delta T}=0$, open with $\nu_{\Delta
T}$ allowed to float.  The coverage of the {\it firs} experiment is
much less extensive and more inhomogeneous over the observed patch
than for {\it dmr}: it was a balloon experiment (taking useful data
for only about 5 hours) with bolometer detectors probing the 4
frequency channels shown in Fig.~\ref{fig:dtspec}. Only the 170 GHz
channel has been fully analyzed. The pixel size, $1.3^\circ$,
and the beam-size, $\ell_s \approx 34$ (a $3.9^\circ$ {\it fwhm} beam)
are half the COBE values. The {\it firs} map has a positive
cross-correlation with {\it dmr} \cite{firsdmr}.  The band-powers of
{\it dmr} and {\it firs} are quite comparable, almost independent of
degree of the signal-to-noise filtering and which frequencies are
probed. And it is only weakly dependent upon the slope $\nu_{\Delta
T}$ (as the formula for $\avrg{{\cal C_\ell}}_{dmr}$ as a function of
$\nu_{\Delta T}$ given by eq.(\ref{eq:sig8cldmr}) shows).

The effective slope of the standard CDM model is $\nu_{\Delta T}
\approx 0.15$ over the {\it dmr} band; variation in $\Omega_B$ and
$H_0$ does not change this very much. {\it dmr} and {\it firs} have
enough coverage in $\ell$-space that one can estimate the spectral
index from the data as well by Bayesian means. If there is no
filtering the 53+90+31 GHz `$A$+$B$' index is $\nu_{\Delta T}$ is
$0.07\pm 0.28$ for {\it dmr}; {\it firs} gives a steeper index,
$0.6^{+0.7}_{-0.8}$, but the one sigma error bar encompasses $0$ and
there is clearly a small angle residual `noise' driving the higher
values  \cite{capripow,bh95}. When CDM models
are marginalized over $\sigma_8$ the preferred {\it dmr} index is 
$n_s=1.02^{+.23}_{-.25}$ for the case with no
gravity waves ($\nu_t=0$), and $n_s=1.02^{+.23}_{-.18}$ with
gravity waves ($\nu_t=\nu_s$) \cite{bj96}. Band-powers for specific
$\ell$ ranges
also show the nearly flat character for ${\cal C}_\ell$ (light
open points at $\ell \sim 4,8,16$ from \cite{dmr4h}). 

The Tenerife point \cite{ten} at $\ell \sim 20$ uses combined 15 and
33 GHz data. The amplitude is compatible with {\it dmr} and {\it firs}, with
CDM-like models, and also have features in common with {\it dmr}.

We now come to the crowded region from two degrees to half a
degree. The next two experiments are the South Pole HEMT experiments of the
UCSB group 
and the Saskatchewan (Big Plate) HEMT experiment of the
Princeton group. The lower open circle is from a joint 4-channel analysis of
the 9 and 13 point {\it sp91} scans \cite{sp91913,india,capripow,bj96}
(with the individual 9 point and 13 point values given in the lower
panel). The upper solid point is for a simultaneous analysis of all channels
of the {\it sp94} data \cite{sp94,bj96}, with separate values for the
Ka ($\sim 30~{\rm GHz}$) and Q ($\sim 40~{\rm GHz}$) HEMT 
bands (Fig.~\ref{fig:dtspec}) in the lower panel. The solid
triangle in the upper panel is the {\it sk93} result \cite{bp}; the big solid
circle at $\ell \sim 80$ in the lower panel is the {\it sk93+94}
result (with calibration uncertainties adding another 14\% error 
to the statistical error shown). 
The nearness of the {\it sp94}, {\it sk93} and {\it sk93+94}
band-powers, and the demonstration for both experiments that the
preferred frequency dependence is nearly flat in $\Delta T/T$ and many
sigma away from bremsstrahlung or synchrotron, the expected
contaminants in this 30-40 GHz range, lend confidence that the
spectrum in the $\ell \sim 60$--$80$ region has really been
determined; and it looks quite compatible with the COBE-normalized CDM
spectrum: {\it sp94} gives $\sigma_8=1.26^{+.37}_{-.27}$, and
{\it sk93+94} gives $\sigma_8=1.21^{+.24}_{-.19}$ \cite{bj96},
very close to the {\it dmr} value $\sigma_8=1.20\pm
0.08$ and the {\it firs} value $\sigma_8=1.27\pm
0.30$.  The 5 heavy open circle points probing
$\ell$'s ranging from 60 to 400 repeated in the upper and lower panels
labelled {\it sk95} are combined {\it sk94+95} results
\cite{sk95}. The estimated 14\% error in the overall amplitude because
of calibration uncertainties are included. The
large $\ell$-space coverage from this one intermediate angle
experiment gives a first glimpse of the $\ell$-space coverage that
will become standard in the next round of anisotropy experiments.

Python \cite{py}, {\it py}, the heavy solid curve at $\ell \sim 90$,
is sensitive to a wide coverage in $\ell$-space as the horizontal
error bars in the top panel indicate. Argo \cite{argo}, {\it ar}, a
balloon-borne experiment, is next. The next five points in the lower
panel are from the fourth and fifth flights, {\it M4,M5}, of the MAX 
balloon experiment
\cite{max5,max4}. Because the filters changed with frequency, the
points are placed at
the average over all {\it max} filters. In the upper panel three {\it
max4} scans are combined into one data point as are two {\it max5}
scans. The lines ending in triangles at $\ell \approx 145$ and 240
denote the 90\% limits for the MSAM \cite{msam} single ({\it msam2})
and double ({\it msam3}) difference configurations. A limitation on
these balloon experiments is the $\sim 5$ hours over which data can be
effectively taken. Planned long duration balloon flights that would
circle Antarctica for about a week would allow extensive mapping at
high precision to be done, and a number of groups have been proposing
designs ({\it e.g.}, ACE, Boomerang and Top Hat).

The CAT points \cite{cat96} at $\ell \approx 400$ and 600 represent a
very different experimental technique, interferometry. CAT is a
3-element synthesis telescope, probing $\sim 15$ GHz frequencies with
a $ 27^\prime$ synthesized beam and a $2^\circ$ field-of-view (the
{\it fwhm} of the individual telescopes). It is a precursor to the
larger VSA (Very Small Array), covering a wider frequency range with
more telescopes and a larger ($4^\circ$) FOV. Two other CMB
interferometers are also planned: CBI and VCA. The {\it ovro}
experiments also probe radio frequencies, but using single dishes. The
historically important 1987 {\it ovro} 7 point upper limit \cite{ov7}
shown used a 40 meter dish.  Detections using as well a 5 meter dish
have now been found with {\it ovro} and give a value in between the 2
CAT points with about the same amplitude. The open triangle at $\ell
\approx 160$ denoting the 95\% credible limit for the {\it sp89} 9
point scan \cite{sp89,belm} was also historically important. WhiteDish
\cite{whitedish} had a small amplitude filter, a hint of a detection
in the $m=1$ mode, and a 95\% limit in $m=2$ mode at $\ell \approx
520$, {\it wd2}.

\section{COBE-normalization of Post-Inflation Fluctuations}\label{pzeta}

 For early universe calculations and also to characterize the initial
conditions for the photon transport through decoupling, the power in
adiabatic scalar fluctuations on scales beyond the Hubble radius is
best characterized in terms of quantities which become
time-independent. Some examples are the spatial curvature of time surfaces
on which there is no net flow of momentum ($\propto \varphi_{com}$), 
as Chibisov and Mukhanov
emphasized long ago \cite{mukphim,bardeen80,LythStewart},
the expansion factor fluctuation, $\delta \ln a\vert_{H*}$, on time
surfaces with uniform space creation rate $H_*$ 
 \cite{sb12,bh95}, and Bardeen, Steinhardt and Turner's $\zeta$  
 \cite{bst,bbks,sbb}. 
An initially scale invariant adiabatic spectrum has
$k$-independent power per $d\ln k$ in these variables (for
$k/(\bar{H}\bar{a}) \ll 1$), while for models with spectral tilt $\nu_s$, 
we have ${\cal P}_{\varphi_{com}} (k) = {\cal P}_{\varphi_{com}} 
(\tau_0^{-1})(k\tau_0)^{\nu_s}$, where we use
the instantaneous comoving horizon size at the current epoch,
$\tau_0$, 
as the normalization point. 
For CDM-like models (those with $\Omega=\Omega_{nr}=1$ and $\tau_0=
2H_0^{-1}$), these are related to
the portion of the {\it dmr} band power $\avrg{{\cal C}_\ell}_{dmr}$
in the scalar adiabatic mode, $\avrg{{\cal C}^{(S)}_\ell}_{dmr} =
\avrg{{\cal C}_\ell}_{dmr}/(1+\tilde{r}_{ts})$, and to the quadrupole
power, ${\cal
C}^{(S)}_2 ={\cal C}_2 /(1+{r}_{ts})$, by \cite{bh95}
\begin{eqnarray}
&& {\cal P}_{\varphi_{com}} (\tau_0^{-1})  \approx 23.5 \avrg{{\cal
C}^{(S)}_\ell}_{dmr} \ e^{-1.99\nu_s (1+0.1\nu_s)} \approx 23.6{\cal
C}^{(S)}_2 \ e^{-1.1\nu_s}\, ,  \label{eq:PzetaCLdmr} \\
&& {\cal P}_{{\varphi}_{com}}(k ) 
\approx 
 {\cal P}_{{\ln a}\vert_{H*}}(k) 
\approx {\cal P}_\zeta (k)  \, \quad {\cal P}_{{\varphi}_{com}}(k ) = {\cal
P}_{\varphi_{com}} (\tau_0^{-1})(k\tau_0)^{\nu_s}\, , \nonumber  
\end{eqnarray}
\ie about $3\times 10^{-9}$. This relation is very insensitive to
variations in ${\rm h}$ and $\Omega_B$.  For scales of order our
present Hubble size, we also have ${\cal P}_\zeta \approx {25\over 9}
{\cal P}_{\Phi_N}\approx {25\over 4} {\cal P}_{(\delta \rho)_{hor}}$,
where $\Phi_N$ is the perturbed Newtonian gravitational potential and
$(\delta \rho)_{hor}$ is the density fluctuation at `horizon
crossing', defined by $k\tau =1$. 

Quantum noise in the transverse traceless modes of the perturbed
metric tensor would also have arisen in the inflation epoch and for
many models may have been quite significant \cite{GWStarob,AW,cbdes}.
The gravitational
radiation power spectrum ${\cal P}_{GW} = {\cal P}_{h_{+}}+{\cal
P}_{h_{\times }}$ is the sum of the two independent gravitational wave
polarizations. It is related to the amplitude of the {\it dmr} band
power $\avrg{{\cal C}^{(T)}_\ell}_{dmr} = \avrg{{\cal
C}_\ell}_{dmr}\tilde{r}_{ts}/(1+\tilde{r}_{ts})$ and to the quadrupole
${\cal C}^{(T)}_2$ by 
\begin{eqnarray}
&& {\cal P}_{GW}(\tau_0^{-1}) \approx  17.6 \avrg{{\cal
C}^{(T)}_\ell}_{dmr} \ e^{-1.92\nu_t (1+0.1\nu_t)}\approx  13.4 {\cal
C}^{(T)}_2 \ e^{-1.25\nu_t}\, ,   \nonumber \\
&& {\cal P}_{GW}(k)= {\cal P}_{GW}(\tau_0^{-1})(k\tau_0)^{\nu_t}
\, . \label{eq:PGWCLdmr}
\end{eqnarray}

The inflation model determines the ratio of ${\cal
P}_{GW}(\tau_0^{-1})$ to ${\cal P}_{{\varphi}_{com}}(\tau_0^{-1})
$. It is generally related to the tilt of the gravity wave spectrum,
and this can in turn by used to relate the ratio of dmr band-powers
(and quadrupoles) to the tilts (for $\Omega_{vac}=\Omega_{curv}=0$):
\begin{eqnarray}
&&{\cal P}_{GW}(\tau_0^{-1})/{\cal
P}_{{\varphi}_{com}}(\tau_0^{-1}) = -4\nu_t /(1-\nu_t/2)\ ; \ \ {\rm
hence} \nonumber \\ &&\tilde{r}_{ts} \equiv {\avrg{{\cal
C}^{(T)}_\ell}_{dmr} \over \avrg{{\cal C}^{(S)}_\ell}_{dmr}} \approx
5.4{(-\nu_t) \over (1-\nu_t/2)} \, e^{-0.07 \nu_t} \, e^{-1.99(\nu_s -
\nu_t )} \, , \nonumber \\ && {r}_{ts} \equiv {{\cal C}^{(T)}_2 \over
{\cal C}^{(S)}_2} \approx 6.9{(-\nu_t )\over (1-\nu_t/2)} \, e^{-0.15
\nu_t} \, e^{-1.1(\nu_s - \nu_t )} \, . \label{eq:PGPzeta} 
\end{eqnarray}
The tensor tilt is simply related to the deceleration parameter
$q=-a\ddot{a}/\dot{a}^2$ of the Universe in the inflationary epoch,
$\nu_t/2 \approx 1+q^{-1}$; although $1+q^{-1}$ is the leading term
for the scalar tilt, other terms can dominate when the deceleration is
near the critical deSitter-space value of $-1$ (\eg \cite{yukawa93}). Thus
although $\nu_t$ is negative, $\nu_s$ may not be. 

When assessing the effect of gravity waves on the normalization of the
spectrum, it is useful to consider two limiting cases: $\nu_s \approx
\nu_t$, which holds for the widest class of models, including power
law and chaotic inflation, and $\nu_t \approx 0$, with $\nu_s$
arbitrary, which holds for some models such as ``natural'' inflation.

There are also corrections as one goes away from the
$\Omega_{nr}=\Omega_{tot} =1$ models. For example, models with nonzero
cosmological constant $\Omega_{vac}$, but
$\Omega_{nr}+\Omega_{vac}=1$, have ${\cal P}_{GW}/\avrg{{\cal
C}^{(T)}_\ell}_{dmr}$ being only weakly dependent upon $\Omega_{vac}$
whereas ${\cal P}_\zeta /\avrg{{\cal C}^{(S)}_\ell}_{dmr} \propto
(1-0.6\Omega_{vac}^{3.5}) $ is strongly dependent upon it.

\section{Using COBE, FIRS, SP94, SK94, ...  to fix $\sigma_8$} \label{sig8}

Before the COBE detection, normalization of the density spectrum was
done using $\sigma_8$, the rms (linear) mass density fluctuations on
the scale of $8\hmpc$, or to a biasing factor $b_g$ for galaxies,
which was usually assumed to obey $b_g \sigma_8 \approx 1$ \eg
\cite{be84,belm}. The relation of
$\sigma_8$ to the initial power spectrum amplitude ${\cal
P}_{{\varphi}_{com}}(\tau_0^{-1})$ is more sensitive to the
specifics of the model, such as type of dark matter, spectral slope,
${\rm h}$, than is the {\it dmr} band-power relation given 
in \S~\ref{pzeta}. Comparing $\sigma_8$ estimates from large
scale structure observations with the COBE-normalized value is thus
extremely important for constraining cosmological parameter space. 
In this section I present a useful functional form, 
$\sigma_8(\Gamma , \nu_s , \nu_t , \Omega_{cdm}, \Omega_{m\nu} ,
\Omega_{vac} )$, where $\Gamma$ parameterizes the density power
spectrum, then use constraints on $\sigma_8$, $\Gamma$ and the tilts
to sketch which models can already be ruled out.

A byproduct of the linear perturbation calculations used to
compute $\Delta T/T$ is the transfer function, 
which maps the initial density fluctuation spectrum in
the very early universe into the final post-recombination one. Many
fits to transfer functions have been given in the literature. One of
the most useful exploits the approximate scaling with the "horizon" at
redshift $\Omega_{nr}/\Omega_{er}$ when the density in nonrelativistic
matter, $\Omega_{nr}\bar{a}^{-3}$, equals that in relativistic matter,
$\Omega_{er}\bar{a}^{-4}$, $k_{Heq}^{-1} = 5 \ \Gamma_{eq}^{-1} \
\hmpc$. The factor $\Gamma_{eq} = \Omega_{nr} \, {\rm h} \
[\Omega_{er}/(1.68\Omega_{\gamma})]^{-1/2}$ provides the main shape
dependence, but a further correction factor can approximately
incorporate the effect of baryons for $\Omega_B \ll \Omega_{nr}$ over
the large scale structure region in $k$-space 
 \cite{peacockdodd94,sugiyama94supp}. 
One functional form for this is \cite{ebw} 
\begin{eqnarray}
&& {\cal P}_\rho (k) \propto  k^{4+\nu_s(k)} \, 
\Big\{1 + [ak + (bk)^{3/2} + (ck)^2]^{p}\Big\}^{-2/p} \, D_\nu^2
\,  , 
\label{eq:tk} \\ 
&&(a,b,c) = (6.4,3.0,1.7) \Gamma^{-1} \hmpc \, , p =1.13 \,  ,
\nonumber \\
&& \Gamma \approx \Omega_{nr} \, {\rm h} \
\Big[{\Omega_{er}\over (1.68\Omega_{\gamma})}\Big]^{-1/2}
\, e^{-(\Omega_B(1+\Omega_{nr}^{-1}(2{\rm h})^{1/2}))
-0.06)}\, .  \nonumber 
\end{eqnarray}  
(Another functional form uses the well known $\Omega_B\rightarrow 0$
BBKS version of the CDM transfer function as a base upon which the
$\Gamma$ variations are imposed.)  
Generally, more scales are needed to characterize the spectrum than
just $k_{Heq}$; \eg the collisionless damping scale for hot dark
matter (massive neutrinos) $k_{\nu damp}^{-1} \approx 6\ (\Omega_{nr}
\Omega_{\nu} (2{\rm h})^2)^{-1/2} (g_{m\nu}/2)^{1/2}\ \hmpc$, with
$g_{m\nu}$ the number of massive species (counting particle and
antiparticle). The correction factor for massive 
neutrinos, $D_\nu$, is fit by \cite{mdmpogosyan}, and is quite
accurate even for finite $\Omega_B$ \cite{lithwick95}. 

Estimations of $\sigma_8$ from the {\it dmr} data for 
selected models can calibrate a scaling relation between
$\sigma_8$ and $\avrg{{\cal C_\ell}}_{dmr}^{1/2}$ 
found using the ``naive'' Sachs Wolfe formula $\Delta T/T \sim \Phi_N/3$
relating temperature fluctuations to gravitational potential
fluctuations \cite{ebw,abffo,yukawa93,bdmr294,bh95}:
\begin{eqnarray}
&&\Gamma\mbox{-law}\colon\hspace{8.6pt} \quad \sigma_8 \approx
{1.25\over f_{SW}} {10^{5} \avrg{{\cal C_\ell}}_{dmr}^{1/2}\over
(1+\tilde{r}_{ts})^{1/2} } \,{\Omega_{nr}^{-0.77} (2(\Gamma -0.03))
\over (1+0.55 (\Omega_{m\nu} /0.3)^{1/2})} \, e^{2.63\nu_s} ,
\label{eq:sig8cldmr} \\ 
&& \mbox{4yr(53\&90\&31)a\&b}\colon \quad 10^5
\avrg{{\cal C_\ell}}_{dmr}^{1/2} \approx [0.82+0.26(1-{\nu_{\Delta
T}\over 2})^{2.8}] \times 1^{+.07}_{-.06} \, , \nonumber \\ 
&& f_{SW}
\approx (1+0.12\Omega_{vac})(1+\Omega_{vac}^{10})\, , \quad
\nu_{\Delta T} \approx 0.15(1-\Omega_{vac})+\nu_s \, , \nonumber \\ 
&&
\tilde{r}_{ts} \approx 5.4{(-\nu_t) \over (1-\nu_t/2)} \, e^{-0.07
\nu_t} \, e^{-1.99(\nu_s - \nu_t )} \, (1-0.6\Omega_{vac}^{3.5})\, .
\nonumber 
\end{eqnarray} 

\begin{figure}
\vspace{-0.7in}
\centerline{\epsfxsize=6.0in\epsfbox{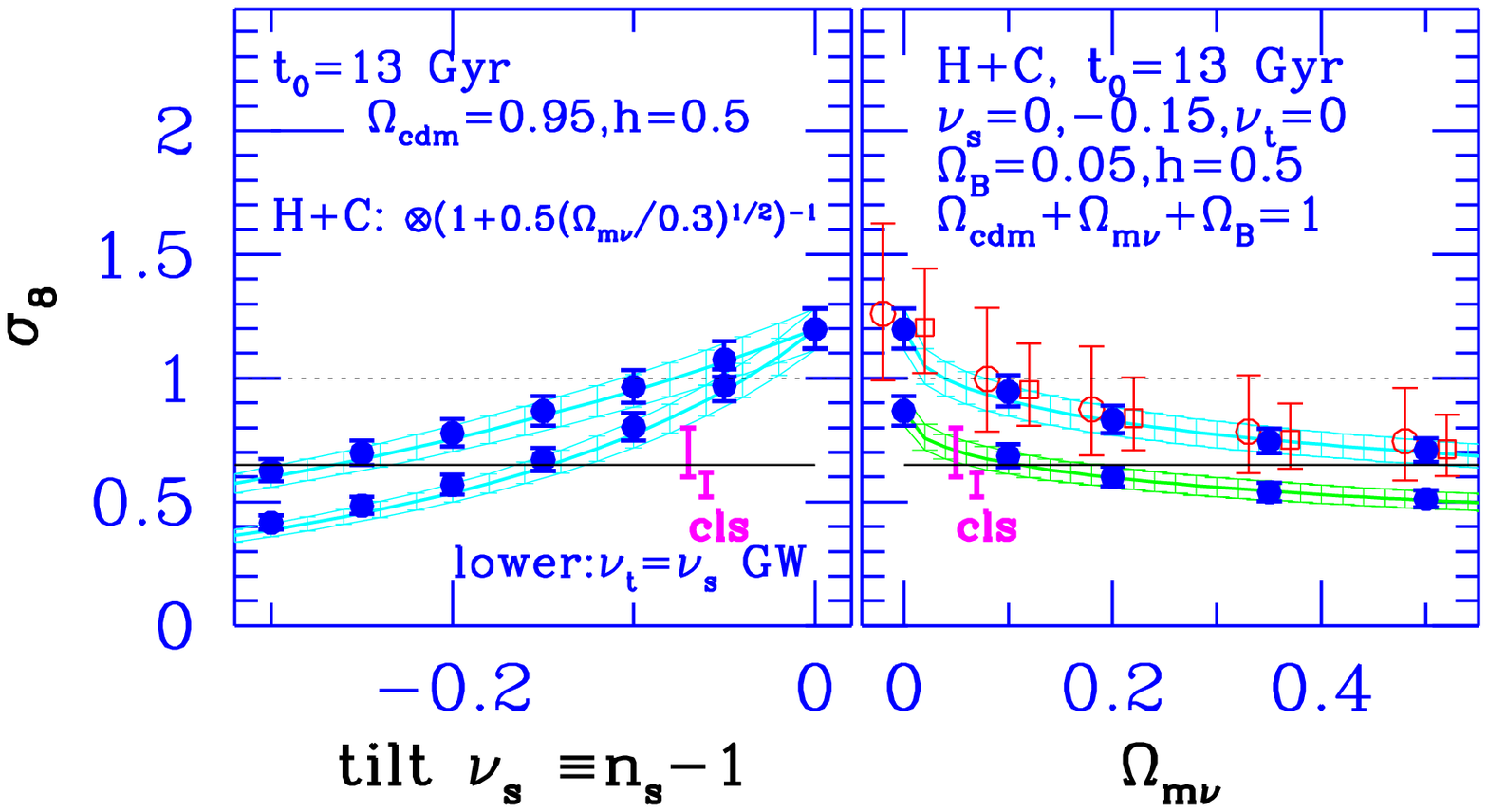}}
\vspace{-3.0in}
\centerline{\epsfxsize=6.0in\epsfbox{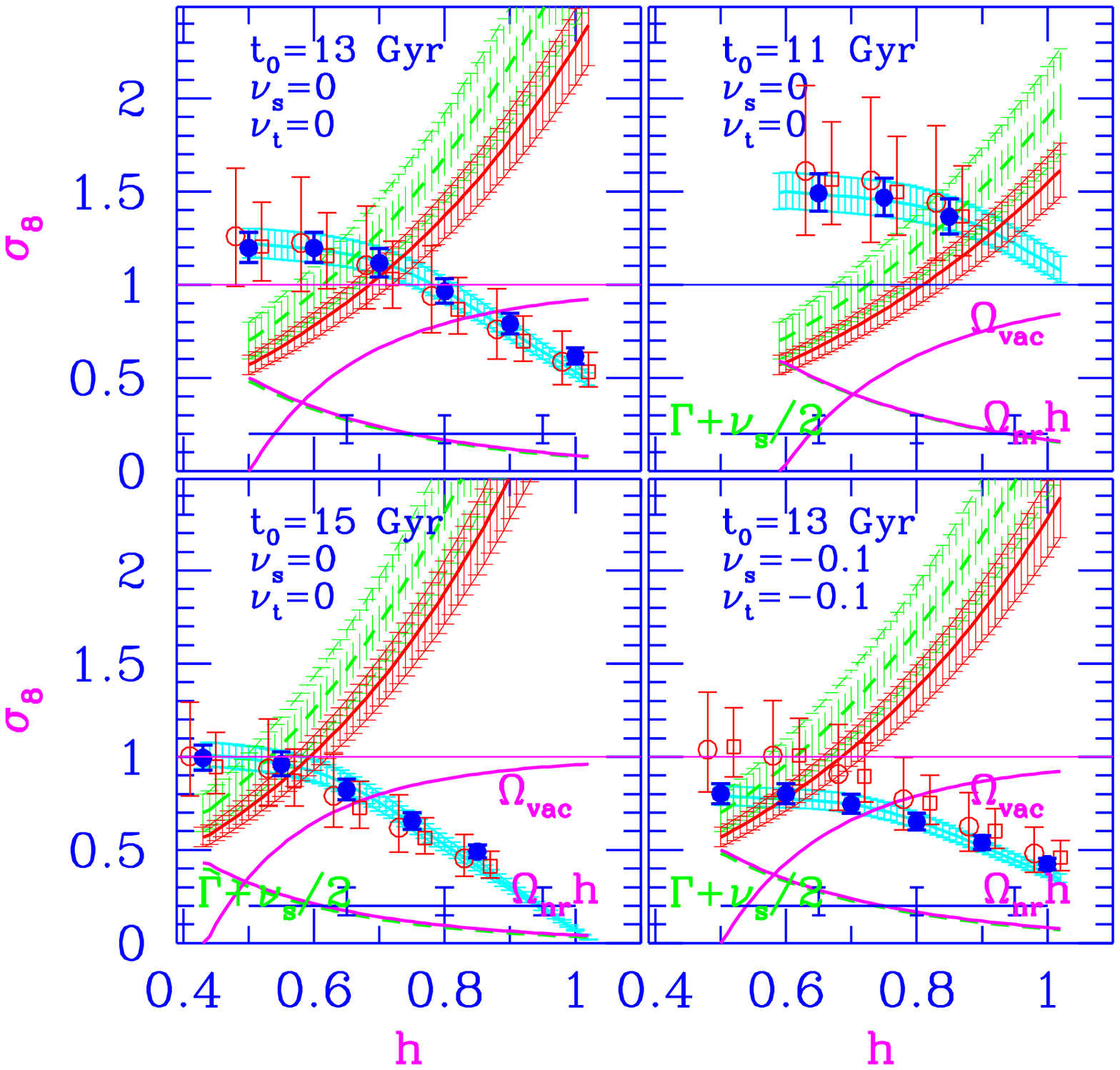}}
\vspace{-0.5in}
\caption{\label{fig:sig8dmr} This illustrates the accuracy and utility
of the fitting formula for $\sigma_8$ for a variety of
inflation-inspired cosmological models. The lower figure also includes
large scale structure and cluster abundance constraints. The symbols
are defined in the text.}
\end{figure}

In {\bf Figure~\ref{fig:sig8dmr}}, the top left panel shows the
average and $\pm 1\sigma$ variation of $\sigma_8$ against tilt for a
pure CDM model when no gravity wave induced anisotropies are included
(upper hatched region) and when they are, for equal tensor and scalar
tilts (lower hatched region). The formula labelled $H+C$ shows how
much $\sigma_8$ is reduced for hot/cold hybrid models (for one species
of massive neutrino, the other two assumed to be effectively
massless).  The heavy closed circles with the error bars denote values
obtained in a Bayesian analysis of the 4-year {\it dmr} (53+90+31)A+B
COBE maps using the exact ${\cal C}_\ell$.  The upper right
panel shows the variation for hot/cold hybrid models with variable
$\Omega_{m\nu}$, for the untilted (upper) case and the $n_s$=0.85
tilted (lower) case (with no gravity wave induced anisotropies).  The
open circles with error bars to the left of the {\it dmr} closed
circles denote $\sigma_8$'s derived from the {\it sp94} data, while
the squares denote $\sigma_8$'s from the {\it sk93+94} data.

The next four panels show $\sigma_8$ against the Hubble parameter
${\rm h}$ for $\Omega_{tot}=1$ models defining a fixed timeline in
parameter space; {\it i.e.}, enough vacuum energy, $\Omega_{vac}
\equiv \Lambda /(8\pi G_N)$ = $1-\Omega_{nr}$, has been added to keep
the total $\Omega$ unity and the age constant.  Using the
$\Omega_{vac}$ dependences of $f_{SW}$, $\nu_{\Delta T}$ and
$\tilde{r}_{ts}$ allow good $\sigma_8$ fits.\footnote{ Although
$f_{SW} =1$ takes into account some of the enhancements over the naive
Sachs--Wolfe formula by normalizing to the calculated
$\sigma_8$--$\avrg{{\cal C_\ell}}_{dmr}^{1/2}$ relation for standard
CDM, it does not take into account the enhancement of $\avrg{{\cal
C_\ell}}_{dmr}^{1/2}$ associated with the time dependence of the
gravitational potential when $\Lambda$ dominates: to take this into
account, we need $f_{SW}$ to exceed unity, as shown.} All models have no mean
curvature, $\Omega_B{\rm h}^2$=0.0125, with the rest of the {\it
nr}-matter in cold dark matter ($\Omega_{nr}=\Omega_{cdm}+\Omega_B$).
For these sequences of models with a uniform age $t_0$, the variation
of $\Omega_{vac}$ with Hubble parameter (the rising curve in
Fig.~\ref{fig:sig8dmr}) is
\begin{eqnarray} 
&& h =h_1 \Omega_{nr}^{-1/2} { \ln [
\sqrt{{\Omega_{vac}/\Omega_{nr}}}+\sqrt{{\Omega_{vac}/\Omega_{nr}}+1}]
\over \sqrt{{\Omega_{vac}/\Omega_{nr}}}} \, , \quad h_1\equiv
0.5(13{\rm Gyr}/t_0) \, , \nonumber \\ 
&& \Omega_{vac} (h) \sim 0.9 (0.3({\rm h}/{\rm h}_1-1)^{0.3}
+ 0.7({\rm h}/{\rm h}_1-1)^{0.4})\, . \label{eq:homv}
\end{eqnarray} 
The latter is a rough inversion. The ages shown in
Fig.~\ref{fig:sig8dmr} bracket a recent Hipparcos-modified estimate
for globular cluster ages, $11.7 \pm 1.4~{\rm Gyr}$
\cite{chaboyerGC97}, with perhaps another Gyr to be added associated
with the delay in globular cluster formation.  The $\Omega_{vac}=0$
model with 13 Gyr age is therefore the $H_0=50$ standard CDM model;
for the 15 Gyr age, $H_0=43$ and for the 11 Gyr age, $H_0=59$.

 The points with error bars again denote the {\it dmr}, {\it
sp94} and {\it sk93+94} values.  The 4-year {\it dmr} data alone
places no useful constraint on $\Omega_{vac}$ or $H_0$: for a scale
invariant spectrum with 13 Gyr age, $\Omega_{vac} < 0.85$ and $H_0 <
88$ at the $1\sigma$ level. The constraint is slightly tighter but
still weak when {\it sp94} and {\it sk93+94} are added: $\Omega_{vac}
< 0.56$ and $H_0 < 65$ at $1\sigma$, $< 0.83$ and $< 85$
at $2\sigma$. For a 15 Gyr age, $\Omega_{vac} < 0.35$ and $H_0 < 51$ at
$1\sigma$, $< 0.83$ and $< 73$ at $2\sigma$. When the slope is
marginalized over, the constraints relax. 
In the figure, the untilted 13 Gyr case is augmented by a tilted case, 
with $n_s =0.9$ and a matching tensor
tilt, $\nu_t=\nu_s=-0.1$. It is not quite flat enough for there to be 
problems matching both the
{\it sk} and {\it dmr} data within the error bars (\eg for $H_0=50$
and 13 Gyr, we get \cite{bj96} $n_s=0.95^{+0.05}_{-0.08}$ with gravity waves,
$0.93^{+0.13}_{-0.12}$ without; for $H_0=70$, $n_s=0.92^{+0.05}_{-0.05}$ with,
$0.88^{+0.14}_{-0.12}$ without; when $H_0$ is marginalized, these numbers
are $0.93^{+0.05}_{-0.07}$ and $0.92^{+0.13}_{-0.14}$.) As is visually
evident from Fig.~\ref{fig:pow}, using the {\it sk95} rather than the {\it
sk94} data with {\it dmr4} increases the preferred $n_s$. However,
when all smaller angle data is used, and $H_0$ is marginalized,
$n_s=1.0 \pm 0.04$ is obtained, a rather encouraging result for
inflation-based models. 

The mass enclosed within $8\hmpc$ is that of a typical rich cluster,
$1.2\times 10^{15} \Omega_{nr}\, (2{\rm h})^{-1} \msun$. Because rich
clusters are rare events in the medium, their number density is
extremely sensitive to the value of $\sigma_8$; the relative numbers
of rich and poor clusters also depends upon the shape of ${\cal
P}_\rho(k)$ in the cluster band, $\sim (0.15-0.25)\, {\rm h}{\rm \, Mpc}^{-1}$, \ie on $n_s$
and $\Gamma$.  Cluster X-ray data implies $0.6 \lta \sigma_8 \lta 0.8$
for CDM-like $\Omega_{nr}=1$ theories, with the best value depending
upon $\Gamma$, $n_s$, some issues of theoretical calibration of
models, and especially which region of the $dn_{cl}/dT_X$ data one
wishes to fit, since the data prefer a local spectral index $d\ln
{\cal P}_{\rho}/d\ln k$ substantially flatter over the cluster region
than the standard CDM model gives \cite{bm3}.  I believe a good target
number is 0.7 and values below about 0.5 are unacceptable, but because
CDM spectra do not fit the data well, this normalization depends upon
whether one focusses on the high or low temperature end. Other authors
who concentrated on the low to median region found lower values for
$n_s=1$ models, $0.57 \pm 0.05$ \cite{wef} and $0.50 \pm 0.04$
\cite{ekecole96}, but do not fit the high $T_X$ end well.\footnote{A small
upward correction should be applied to these low estimates to account
for the nonzero redshift of the calibrating samples.}  In the top left
and right panels, the two vertical lines denote two estimates of
$\sigma_8$ from clusters for $\Omega_{vac}$=0 (which depends somewhat
upon tilt).  For $\Omega_\Lambda > 0$ models, higher values are needed
to fit the cluster data \cite{ebw,kofman93}; \cite{wef} adopt
$\Omega_{nr}^{-0.56}$ as the correction, \cite{ekecole96} give a more
moderate dependence, $\Omega_{nr}^{-0.53+0.13\Omega_{nr}}$.  The
rising curves with error bars in Fig.~\ref{fig:sig8dmr},  using the 
$\Omega_{nr}^{-0.56}$ scaling, show the
higher and lower $\sigma_8$ estimates. Allowed models
would have to lie in the overlap region between the cluster $\sigma_8$
and the {\it dmr} $\sigma_8$. 

  The upper rising regions also roughly
denote the $\sigma_8$ behaviour as derived from optical galaxy
samples, in units of $b_g \sigma_8$ where $b_g$ is the biasing factor
for galaxies.\footnote{There are many estimates of the combination
$\sigma_8 \Omega_{nr}^{0.56}$ that are obtained by relating the galaxy
flow field to the galaxy density field inferred from redshift surveys,
which all take the form $\sigma_8 \Omega_{nr}^{0.56} = [b_g\sigma_8]
\beta_g$, where $\beta_g$ is a numerical factor whose value depends
upon data set and analysis procedure. Strauss and Willick 
 \cite{strausswillick} have reviewed 
the rather varied estimations and give raw averages: $0.78\pm 0.33$
for IRAS-selected galaxy surveys, $0.71\pm 0.25$ for
optically-selected galaxy surveys. Even more recent estimates give
$\beta_g\sim 0.4$--$0.6$.  These $\sigma_8$ estimates rely on
the simplifying assumption of a linear amplification bias $b_g$
for galaxies.  The traditional estimate is  
$b_g=\sigma_8^{-1}$, but $b_g$ can depend upon the
galaxy types being probed, upon scale, and could be bigger or smaller
than $\sigma_8^{-1}$, and certainly cannot be determined by theory
alone.  A recent estimate using the Mark III velocity data is 
$\sigma_8\Omega_{nr}^{0.56} \sim 0.85 \pm 0.1$, with
sampling errors adding another $\sim 0.1$ uncertainty
\cite{zaroubi96,kolattdekel96}. (The $\Omega_{nr}^{0.56}$ 
is the factor by which the linear
growth rate $\dot{D}/D$ differs from the Hubble expansion rate
$\dot{a}/a$.)}  How seriously we take the constraints derived 
using flows and redshift
surveys depends upon how reliable we think the indicators are -- a
subject of much debate. Much work has also gone into relating bulk
flows of galaxies to CMB anisotropy observations on intermediate
angular scales, since they probe the same bands in $k$-space, and this
has led to significant constraints, but is also subject to much
debate. 
 
The shape of the density power spectrum is almost as powerful a
constraint as $\sigma_8$ is.  To fit the large scale galaxy clustering
data, in particular the APM angular correlation function, requires
$0.15 \simlt \Gamma +\nu_s/2 \simlt 0.3$ \cite{ebw,abffo,bh95},
assuming the galaxy power spectrum on large scales is a linear
amplification of the (linear) density power spectrum.  In the lower
panels of Fig.~\ref{fig:sig8dmr}, the dropping curve shows
$\Omega_{nr}{\rm h}$ and the (almost indistinguishable) dashed one
shows $\Gamma$; the target range is shown by the line at 0.2 with a
few error bars. To lower $\Gamma +\nu_s/2$ into the 0.15 to 0.3 range
one can \cite{bbe,sbb,abffo}: tilt the spectrum (expected at some
level in inflation models); lower ${\rm h}$; lower $\Omega_{nr}$; or
raise $\Omega_{er}$ ($=1.68\Omega_{\gamma}$ with the canonical three
massless neutrino species present). Raising $\Omega_B$ also helps. Low
density CDM models in a spatially flat universe ({\it i.e.}  with
$\Lambda > 0$) lower $\Omega_{nr}$ to $1-\Omega_{\Lambda}$. CDM models
with decaying neutrinos raise $\Omega_{er}$ \cite{bbe,be91}: $\Gamma
\approx 1.08 \Omega_{nr} {\rm h}(1 + 0.96 (m_\nu \tau_d /{\rm kev\,
yr})^{2/3})^{-1/2}$, where $m_\nu$ is the neutrino mass and $\tau_d$
is its lifetime. (Decaying neutrino models have the added feature of a
bump in the power at subgalactic scales to ensure early galaxy
formation, a consequence of the large effective $\Omega_{nr}$ of the
neutrinos before they decayed.)

Although Fig.~\ref{fig:sig8dmr} gives a visual impression of which
models are preferred, this can be put on a more quantitative basis.
Fig.~\ref{fig:sig8dmr} shows $\sigma_8$ is a sensitive function of
$n_s$: for CDM models with $\Omega_{nr}=1$, it is far too high at
$1.2$ for $n_s$=1, but too low by $n_s\approx 0.76$ with the
``standard'' gravity wave contribution ($\nu_t=\nu_s$) or by
$n_s\approx 0.60$ if there is no tensor mode contribution. However,
the shape constraint wants lower $n_s$. In \cite{bj96}, we marginalize
likelihood functions determined with the COBE data (and smaller angle
data) using a prior probability requiring that $\Gamma +\nu_s /2$ be
$0.22\pm 0.08$ and $\sigma_8 \Omega_{nr}^{0.56} $ be
$0.65^{+0.15}_{-0.08}$ in order to condense the tendencies evident in
Fig.~\ref{fig:sig8dmr} into single numbers with error bars.  Threading
the ``eye of the needle'' this way is so exacting that the error bars
are too small to take too seriously. Sample numbers using only the
4-year {\it dmr} data and these priors for the 13 Gyr case are:
$n_s=0.76^{+.03}_{-.03}$ for ${\rm h}=0.5$ with gravity waves,
$0.61^{+.04}_{-.04}$ without, \ie much tilt; for ${\rm h}=0.7$ and
$\Omega_{vac} =0.66$, we get $0.99^{+.03}_{-.02}$ with and
$0.94^{+.04}_{-.04}$ without, \ie little tilt; and when Hubble
parameters in the range from 0.5 to 1 are marginalized over, the
preferred index is $n_s=0.99^{+.06}_{-.04}$ with gravity waves,
$0.95^{+.09}_{-.10}$ without, \ie again little tilt. Combining 
{\it sp94} and {\it sk93+94} with {\it dmr} data does not change the
values nor the error bars by much (it flattens a bit). When all
smaller angle data is used plus LSS and {\it dmr4} data, and $H_0$ is
marginalized, $n_s=0.96 \pm 0.03$ is obtained, again not very
different from the  LSS + {\it dmr4} only result. In both cases, a
nonzero $\Omega_{vac}$ is preferred. 

 For the decaying neutrino model
with $n_s=1$ to have $\sigma_8>0.5$ we need $\Gamma >0.22$, \ie $m_\nu
\tau_d < 14 ~{\rm keV~yr}$. The hot/cold hybrid model formula in
eq.~(\ref{eq:sig8cldmr}) is for one massive neutrino species. As
Fig.~\ref{fig:sig8dmr} shows, an $n_s$=1 ${\rm h}$=0.5 hot/cold 
hybrid model with
$\Omega_\nu < 0.3$ would have $\sigma_8 > 0.8$; however, even with a
modest tilt to $n_s=0.95$ this can drop to 0.7 for $\Omega_\nu =
0.25$. (See also ref.  \cite{mdmpogosyan}.) That is, little tilt 
is required, in contrast to the CDM case. 

It is also evident from Fig.~\ref{fig:sig8dmr} that the cluster data
in combination with the {\it dmr} data stops ${\rm h}$ from becoming
too high for a fixed age, but also would prefer a nonzero $\Lambda$
value, with $H_0\sim 60-70$ for 13 Gyr, and $H_0\sim 50-60$ for 15
Gyr. When the tilt is allowed to vary as well, the preferred values
lower to very near 50 and 43, respectively, \ie with little
$\Omega_{vac}$: $h<0.70$ at $2\sigma$ with gravity waves, $h<0.56$
with no gravity waves for 13 Gyr; $h<0.56$ at $2\sigma$ with gravity
waves for 15 Gyr. For the hot/cold models, the values near 50
and 43 are preferred even more, even with very little tilt.

For open CDM models, the COBE-determined $\sigma_8$ goes down with
decreasing $\Omega$ (and increasing ${\rm h}$). These models are not
so attractive because $\Omega$ drops so precipitously with increasing
${\rm h}$ for fixed age (\eg for 13 Gyr and $H_0=70$,
$\Omega_{tot}$=0.055). Equation~(\ref{eq:sig8cldmr}) has not been
modified to treat open models (see \eg \cite{dmr2sig8gorskiopen}).

Texture and string models require a low $\sigma_8$, $< 0.5$. At this
stage, it is still unclear how much of a problem this is since
non-Gaussian effects can modify the cluster distribution, partly
compensating for the low $\sigma_8$.

 Going to higher $k$-bands, COBE-normalized spectra imply values for
the ``the redshift of galaxy formation'', of quasar formation, and the
amount of gas in damped Lyman alpha systems \etc and this also
significantly restricts the parameter space. However, it is more
dependent on the role gasdynamics may play in defining the objects.

If we are so bold as to assume that we now know the shape of the
density power spectrum over the large scale structure band, and the
amplitude of the power spectrum on cluster-scales, then, in
conjunction with the COBE anisotropy level, the range of inflation and
dark matter models is considerably restricted.  Whether the solution
will be a simple variant on the CDM+inflation theme \cite{bbe},
involving slight tilt (or more radical broken scale invariance),
stable ev-mass neutrinos, decaying ($>${\rm kev})-neutrinos, vacuum
energy, low $H_0$, high baryon fraction, mean curvature, or some
combination, is still open, but can be decided as the observations
tighten, and, in particular, as the noise in the ${\cal C}_\ell$
figure subsides, revealing the details of the Doppler peaks, a very
happy future for those of us who wish to peer into the mechanism by
which structure was generated in the Universe.

\noindent
{\bf ACKNOWLEDGMENTS:} The CMB field has seen an explosion of papers
since the COBE discovery. The reference list given here is
truncated. For a more complete list see
\cite{bh95,SilkScottWhiteannrev}. Thanks to my collaborators Andrew
Jaffe, Yoram Lithwick, and Tarun Souradeep for permission to quote
some of our unpublished work.  Support from NSERC and a Canadian
Institute for Advanced Research Fellowship is gratefully acknowledged.

% FINAL REFERENCING ***********************

%\def\refindent{\par\noindent\hangindent=3pc\hangafter=1 }
\def\preprint#1#2#3{\refindent #1: #2, {\it #3}\par}
\def\article#1#2#3#4#5{\noindent #1: #2, {\it #3\/}{\bf #4}, #5\par}
\def\prd{{\it Phys.~Rev.~D\,}}
\def\prl{{\it Phys.~Rev.~Lett.\,}}
\def\apj{{\it Ap.~J.\,}}
\def\apjl{{\it Ap.~J.~Lett.\,}}
\def\apjsuppl{{\it Ap.~J.~Supp.\,}}
\def\mnras{{\it M.N.R.A.S.\,}}

\end{document}